\documentclass[11pt,letterpaper]{article}
\usepackage{fullpage}

\usepackage{amsmath,amssymb}
\date{}

\usepackage{tikz}
\usetikzlibrary{shapes,arrows,fit,backgrounds, calc,
  decorations.pathmorphing,decorations.markings, backgrounds, arrows.meta}

\tikzset{
Smiley/.style={minimum size = 0.5cm, text width = 0.5cm, text centered},
box/.style= {draw, minimum width = 2.5cm, minimum height =0.5cm, fill
= white},
medbox/.style={draw, rectangle, inner ysep = 1.5em, inner xsep = 0.2cm, fill=colorc!5,
    label={[align=right,shift={(0,-4.5ex)}]north:#1}},
bigbox/.style={draw, rectangle, inner sep = 1.5em, fill=colorb!5,
    label={[align=right,shift={(0,-4.5ex)}]north :#1}},
tight/.style={minimum width = 0cm, minimum height =0cm, inner sep = 0},
thickpath/.style = {draw, ->, thick},
decay/.style = {draw, -{>[scale=4]}, decorate, , decoration = {snake, amplitude = 0.8mm, segment length=4mm, post length = 2mm}}
}
\pgfdeclarelayer{background}
\pgfdeclarelayer{foreground}
\pgfsetlayers{background,main,foreground}

\definecolor{highlight}{RGB}{0,115,179}
\definecolor{colora}{RGB}{0,115,179}
\definecolor{colorb}{RGB}{230,154,0} 
\definecolor{colorc}{RGB}{0,154,128} 
\definecolor{colord}{RGB}{205,10,179}
\definecolor{colore}{RGB}{255,32,0}
\definecolor{colorf}{RGB}{240,228,66}
\definecolor{colorg}{RGB}{90,179,230}
\definecolor{colorh}{RGB}{205,154,179}

\definecolor{colorS}{RGB}{0,154,128}
\definecolor{colorI}{RGB}{255,32,0}
\definecolor{colorR}{RGB}{205,10,179}

\usepackage{color}

\newcommand{\ave}[1]{\left \langle #1 \right \rangle}
\newcommand\T{\rule{0pt}{2.6ex}}
\newcommand\B{\rule[-1.2ex]{0pt}{0pt}}

\usepackage{todonotes}
\usepackage{pgf}

\newcommand{\axes}{\draw [<-,ultra thick] (0,\ymax) node (yaxis) [left] {$t$} |- (2.5,0) node (xaxis) [] {}}
\newcommand{\person}[2]{\draw [thin, -latex] (#1,0) node () [below] {#2} -- (#1,\ymax)}
\newcommand{\partnership}[5]{\fill [gray, opacity=#5] (#1,#3) -- (#2,#3) -- (#2,#4) -- (#1,#4) -- cycle}
\newcommand{\ymax}{2}
\newcommand{\xB}{0.2}
\newcommand{\xA}{1.2}
\newcommand{\xC}{2.2}
\newcommand{\SuccessTrans}[3]{\draw [-, color = red, thick] (#1,#3) -- (#2,#3)}
\newcommand{\FailTrans}[3]{\draw [-, color = black, dashed] (#1,#3) -- (#2,#3)}
\newcommand{\InfectionStart}[2]{\draw [-latex, ultra thick, color = red] (#1,#2) -- (#1,\ymax)}

\newcommand{\pd}[2]{\frac{\partial #1}{\partial #2}}
\title{Saturation Effects and the Concurrency Hypothesis: Insights from an Analytic Model}
\author{Joel C. Miller\footnote{Institute for Disease Modeling,
    Bellevue, WA, USA}~ and 
Anja C. Slim\footnote{School of Mathematical Sciences
 \& School of Earth, Atmosphere, and the Environment, Monash
University, Clayton, VIC, Australia
}}

\begin{document}
\maketitle

\begin{abstract}
Sexual partnerships that overlap in time (concurrent relationships) may play a significant role in the HIV epidemic, but the precise effect is unclear.  We derive edge-based compartmental models of disease spread in idealized dynamic populations with and without concurrency to allow for an investigation of its effects.  Our models assume that partnerships change in time and individuals enter and leave the at-risk population.  Infected individuals transmit at a constant per-partnership rate to their susceptible partners.  In our idealized populations we find regions of parameter space where the existence of concurrent partnerships leads to substantially faster growth and higher equilibrium levels, but also regions in which the existence of concurrent partnerships has very little impact on the growth or the equilibrium.  Additionally we find mixed regimes in which concurrency significantly increases the early growth, but has little effect on the ultimate equilibrium level.  Guided by model predictions, we discuss general conditions under which concurrent relationships would be expected to have large or small effects in real-world settings.  Our observation that the impact of concurrency saturates suggests that concurrency-reducing interventions may be most effective in populations with low to moderate concurrency.
\end{abstract}

\section*{Introduction}

The HIV epidemic has had a significant impact worldwide, but especially so in sub-Saharan Africa~\cite{murray2014global}.  The reasons for this difference are many, complex, and not fully understood~\cite{buve2002spread,kalipeni2004hiv}.  One proposed factor is a greater frequency of sexual partnerships that overlap in time, the so-called ``concurrency hypothesis''~\cite{morris:concurrentnetwork,morris:concurrent}.  This hypothesis has received significant attention, but it is highly controversial (see for example~\cite{mah:concurrent,lurie:concurrent,mah:evidence,morris:barking,epstein:concurrent,lurie:concurrent2,kretzschmar:concurrent} and~\cite{sawers2010concurrent,epstein2011concurrent,goodreau2011decade}).

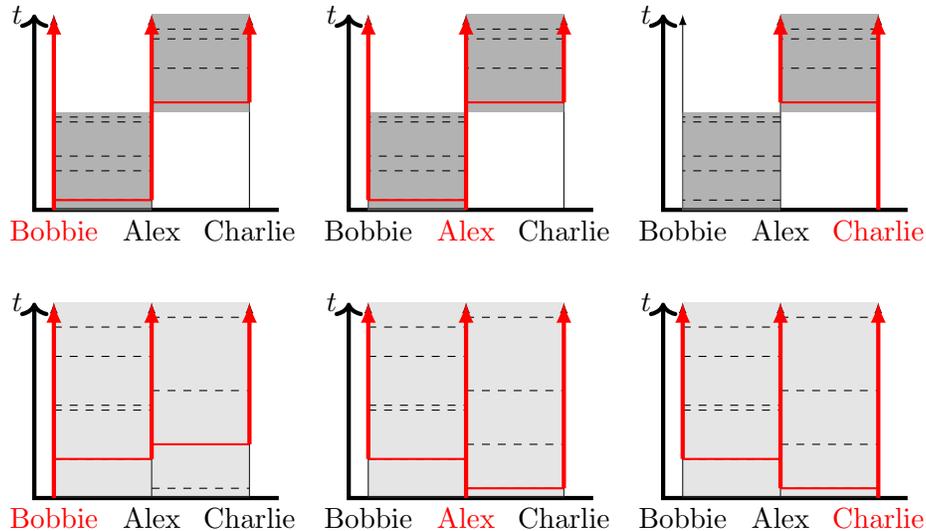
\begin{figure}
\begin{center}
\begin{tikzpicture}[scale=1.3]
\axes;
\person{\xB}{{\color{red}Bobbie}};
\person{\xA}{Alex};
\person{\xC}{Charlie};

\partnership{\xB}{\xA}{0}{0.5*\ymax}{0.6};
\partnership{\xA}{\xC}{1}{\ymax}{0.6};

\foreach \y in {0.1, 0.4, 0.55, 0.9, 0.95}   
   {\FailTrans{\xB}{\xA}{\y};};
\foreach \y in {1.1, 1.45, 1.75, 1.85}
   {\FailTrans{\xA}{\xC}{\y};};

\InfectionStart{\xB}{0};
\SuccessTrans{\xB}{\xA}{0.1}; 
\InfectionStart{\xA}{0.1};
\SuccessTrans{\xA}{\xC}{1.1};
\InfectionStart{\xC}{1.1}; 
\end{tikzpicture}
\begin{tikzpicture}[scale=1.3]
\axes;
\person{\xB}{Bobbie};
\person{\xA}{{\color{red}Alex}};
\person{\xC}{Charlie};

\partnership{\xB}{\xA}{0}{0.5*\ymax}{0.6};
\partnership{\xA}{\xC}{1}{\ymax}{0.6};

\foreach \y in {0.1, 0.4, 0.55, 0.9, 0.95}   
   {\FailTrans{\xA}{\xB}{\y};};
\foreach \y in {1.1, 1.45, 1.75, 1.85}
   {\FailTrans{\xC}{\xA}{\y};};

\InfectionStart{\xA}{0};
\SuccessTrans{\xB}{\xA}{0.1}; 
\InfectionStart{\xB}{0.1};
\SuccessTrans{\xA}{\xC}{1.1};
\InfectionStart{\xC}{1.1}; 
\end{tikzpicture}
\begin{tikzpicture}[scale=1.3]
\axes;
\person{\xB}{Bobbie};
\person{\xA}{Alex};
\person{\xC}{{\color{red}Charlie}};

\partnership{\xB}{\xA}{0}{0.5*\ymax}{0.6};
\partnership{\xA}{\xC}{1}{\ymax}{0.6};

\foreach \y in {0.1, 0.4, 0.55, 0.9, 0.95}   
   {\FailTrans{\xA}{\xB}{\y};};
\foreach \y in {1.1, 1.45, 1.75, 1.85}
   {\FailTrans{\xC}{\xA}{\y};};

\InfectionStart{\xC}{0};
\SuccessTrans{\xC}{\xA}{1.1};
\InfectionStart{\xA}{1.1}; 
\end{tikzpicture}\\[12pt]


\begin{tikzpicture}[scale=1.3]
\axes;
\person{\xB}{{\color{red}Bobbie}};
\person{\xA}{Alex};
\person{\xC}{Charlie};

\partnership{\xB}{\xA}{0}{\ymax}{0.2};
\partnership{\xA}{\xC}{0}{\ymax}{0.2};

\foreach \y in {0.4, 0.9, 0.95, 1.45, 1.75}   
   {\FailTrans{\xB}{\xA}{\y};};
\foreach \y in {0.1, 0.55, 1.1, 1.85}
   {\FailTrans{\xA}{\xC}{\y};};

\InfectionStart{\xB}{0};
\SuccessTrans{\xB}{\xA}{0.4}; 
\InfectionStart{\xA}{0.4};
\SuccessTrans{\xA}{\xC}{0.55};
\InfectionStart{\xC}{0.55}; 
\end{tikzpicture}
\begin{tikzpicture}[scale=1.3]
\axes;
\person{\xB}{Bobbie};
\person{\xA}{{\color{red}Alex}};
\person{\xC}{Charlie};

\partnership{\xA}{\xB}{0}{\ymax}{0.2};
\partnership{\xC}{\xA}{0}{\ymax}{0.2};

\foreach \y in {0.4, 0.9, 0.95, 1.45, 1.75}   
   {\FailTrans{\xB}{\xA}{\y};};
\foreach \y in {0.1, 0.55, 1.1, 1.85}
   {\FailTrans{\xA}{\xC}{\y};};

\InfectionStart{\xA}{0};
\SuccessTrans{\xC}{\xA}{0.1}; 
\InfectionStart{\xC}{0.1};
\SuccessTrans{\xA}{\xB}{0.4};
\InfectionStart{\xB}{0.4}; 

\end{tikzpicture}
\begin{tikzpicture}[scale=1.3]
\axes;
\person{\xB}{Bobbie};
\person{\xA}{Alex};
\person{\xC}{{\color{red}Charlie}};

\partnership{\xA}{\xB}{0}{\ymax}{0.2};
\partnership{\xC}{\xA}{0}{\ymax}{0.2};

\foreach \y in {0.4, 0.9, 0.95, 1.45, 1.75}   
   {\FailTrans{\xB}{\xA}{\y};};
\foreach \y in {0.1, 0.55, 1.1, 1.85}
   {\FailTrans{\xA}{\xC}{\y};};

\InfectionStart{\xC}{0};
\SuccessTrans{\xC}{\xA}{0.1}; 
\InfectionStart{\xA}{0.1};
\SuccessTrans{\xA}{\xB}{0.4};
\InfectionStart{\xB}{0.4}; 

\end{tikzpicture}
\end{center}
\caption{\textbf{Sample scenarios comparing serially monogamous (top) and concurrent (bottom) relationships:}  Shaded regions denote the existence of a partnership between ``Alex'' and either ``Bobbie'' or ``Charlie'', with darker shading representing a partnership having a higher transmission rate.  Dashed lines denote transmission events within the relationship that would cause infection if  one individual were infected and the other susceptible.  Vertical red lines denote time at which an individual is infected, and horizontal red lines denote successful transmissions.  In the concurrent case, the transmission events occur at exactly the same times, but the transmission could occur in either partnership.  Thus the interaction rate within each partnership is half that of the serial case.  Concurrency provides  additional transmission routes and
  tends to speed up onwards transmission.  In the left panels Bobbie begins
  infected, in the cenral panels Alex begins infected, and in the right panels Charlie begins infected. }
\label{fig:concurrency_display}
\end{figure}

\subsection*{Mechanisms of concurrency}
It is worth exploring a simplified scenario to illustrate the key mechanisms by which concurrency can affect disease transmission as well as some of the subtleties, which make ultimate impact less obvious and make observational study design difficult.  Consider an individual ``Alex'' who has two partners ``Bobbie'' and ``Charlie'' over a period of one year.  Two potential partnership arrangements are shown in Fig.~\ref{fig:concurrency_display}.  In the serial case (top of Fig.~\ref{fig:concurrency_display}), Alex's partnership with Bobbie lasts for six months and is replaced by a six-month partnership with Charlie.  In the concurrent case (bottom of Fig.~\ref{fig:concurrency_display}), Alex's partnerships with Bobbie and Charlie overlap completely.  
During each partnership, there are occasional ``transmission events'', or interactions between the individuals that would cause infection if one individual were infected and the other susceptible.  If a transmission event occurs between an infected and a susceptible individual, then it is ``successful'' and the susceptible individual becomes infected.  In Fig.~\ref{fig:concurrency_display}, and throughout this paper, we assume that the rate of potentially infectious interactions per individual is the same.  Thus in a population of only serial partnerships, the rate of interactions per partnership is twice that in a population with exactly two overlapping partnerships per individual. The partnership duration is scaled so that the expected number of transmission events per partnership is the same.  This allows us to ensure that in our comparisons the only change is the level of concurrency, and our results are not conflated with the effect of increased interactions per individual or per partnership.  This is illustrated in Fig.~\ref{fig:concurrency_display} by halving the temporal density of transmission events and doubling the partnership duration in the concurrent case.

First consider direct transmission to Alex from Bobbie or Charlie.  Regardless of the partnership arrangement or which partner is initially infected, the overall risk of infection to Alex is the same.  However, the timing of infection differs.  If Bobbie is initially infected, then Alex tends to be infected earlier in a serial partnership than a concurrent one (because of the focused relationship).  Conversely, if Charlie is initially infected, then Alex becomes infected earlier in a concurrent partnership (because of the delay forced by partnership timing in the serial case).

Now consider indirect transmission between Bobbie and Charlie via Alex.  Significant differences in both the risk and timing of infection now exist depending on partnership arrangement.  If Charlie is initially infected, then a transmission chain from Charlie to Alex to Bobbie is only possible in the concurrent case.  If Bobbie is initially infected, then a transmission chain from Bobbie to Alex to Charlie will happen faster in the concurrent case because there is no built-in delay for partnership change.  In turn this would allow Charlie to begin transmitting to other partners earlier.  However, the probability of a Bobbie to Alex to Charlie transmission chain is slightly reduced because some of the interactions between Alex and Charlie will already have happened by the time Alex is infected (see Fig.~\ref{fig:concurrency_display2} later). 

Finally, consider the result if Alex enters the partnerships already
infected.  The outcomes are the same regardless of whether the
populations have concurrency, but concurrency tends to reduce the
average time to transmission because the partnerships can start
sooner.  Unless Bobbie or Charlie also have concurrent relationships,
this has no population-scale impact.

Thus we anticipate that concurrency will increase the spread of
disease through two key mechanisms: by allowing the disease to trace
transmission routes faster and by providing additional transmission
routes.  The subtleties described above can limit the extent of its
effect.  Whether one individual has concurrent
relationships only affects the outcome of a particular partnership if
both partners are susceptible at the beginning of their partnership
and at least one becomes infected during it.  Furthermore, the
identical risk of infection to Alex regardless of partnership
arrangement in Fig.~\ref{fig:concurrency_display} illustrates that the
risk of concurrency is to partners of the individual with concurrent
relationships rather than to the individual with concurrent relationships.

\subsection*{Modeling approaches}
Unfortunately, measuring or predicting the magnitude of the impact of
concurrency has been difficult.  Concurrency is difficult to directly
measure.  Even when it is identified, observational studies comparing
an individual with serial relationships to an individual with
overlapping relationships within a given sample population will not
test for the effects of concurrency.  Instead the study would need to
compare their partners, which is more difficult.  

Modeling studies have a different set of challenges.  Models are usually either stochastic agent-based simulation [which we will call ``stochastic simulations'' or simply ``simulations''] or equation-based [which we will call ``analytic models''].  Stochastic simulation of concurrency is often difficult because of inherent difficulties in identifying which of many parameters governs an outcome as well as computational limitations on the populations considered.  Analytic models in contrast have difficulties because the standard well-mixed population assumption of analytic models precludes the existence of concurrent relationships. There is a need for analytic modeling that avoids this assumption.

Because analytic models have not existed for populations with concurrent relationships, most modeling investigations of concurrency have used stochastic simulation.  Many are reviewed in~\cite{epstein2011concurrent}.  

Recent work on analytic models has shown how they can be used to
incorporate some partnership structure~\cite{pastor2015epidemic,wang2017unification,kiss:EoN}.   The work of~\cite{leung:demographic,leung:disease, leung2015concurrency} in particular used analytic models to investigate how concurrency alters the epidemic threshold.  The very recent work of~\cite{leung2016dangerous} provides a renewal equation from which dynamics of a model (similar to the one we present below) can be calculated.  The model of~\cite{kamp2010untangling} also provides a dynamic prediction, but it makes simplifying assumptions about the independence of partner status, partnership age, and partner age.  Where correlations build up this can lead to erroneous predictions.

Recent
work~\cite{volz:cts_time,volz:dynamic_network,miller:volz,miller:ebcm_overview,miller:ebcm_structure}
developed an  ``Edge-Based Compartmental Modeling'' (EBCM) approach which
leads to differential equation models with only a handful of
equations.  The models exactly predict the large population dynamics of SI and SIR
disease spread in static random networks~\cite{decreusefond:volz_limit,janson:SIRproof}.  
The approach has been generalized to networks with changing partnerships, but still assuming a closed population~\cite{miller:ebcm_overview}.  Because the HIV epidemic developed over decades, it is not appropriate to ignore individuals entering and leaving the at-risk population as they age.  Further, if we ignore ``birth'' and ``death'' for an SI disease in a closed population eventually the disease reaches the full population.  Nevertheless, this provides a starting point from which we might be able to develop a tractable model that does capture ``birth'' and ``death''.

We adapt the EBCM approach to accommodate ``births'' and ``deaths'' representing entry into and exit from the at-risk population.  We show that the resulting equations accurately predict the outcome of simulations in the large population limit, and our primary focus is on using the model to investigate the role that concurrency can play in the spread of a ``Susceptible--Infected'' (SI) disease such as HIV.  

Our goals in this paper are:
\begin{enumerate}
\item To demonstrate an analytic model for disease spread in a dynamic population with concurrency and
show that it exactly predicts simulated dynamics in the large
population limit.
\item To use the model to identify important regimes under which
  concurrency does or does not have an important effect and to
  understand the underlying mechanism by which this occurs in the model.
\item To explore what features of these underlying mechanisms would
  need to be preserved in real-world scenarios in order for
  concurrency to have (or not have) a major impact on population-scale
  outcomes.
\end{enumerate}
This paper is not intended to be an authoritative statement about the
role of concurrent relationships in Africa, rather we hope that an
improved understanding of the underlying mechanisms  will improve the quality of the discussion.

\section*{Materials and Methods}

In this section we introduce our stochastic population and disease model, state the governing equations for the large population limit, and briefly outline their derivation.  The full derivation is given in the Supporting Information (SI).

\subsection*{Population/Disease Assumptions}
We assume discrete time.  This assumption is made to simplify the
simulations we use to validate the analytic model.  The
continuous time version of the analytic model is presented in the SI.

\begin{figure}
\begin{center}
\raisebox{-0.5\height}{\includegraphics[width=0.3\textwidth]{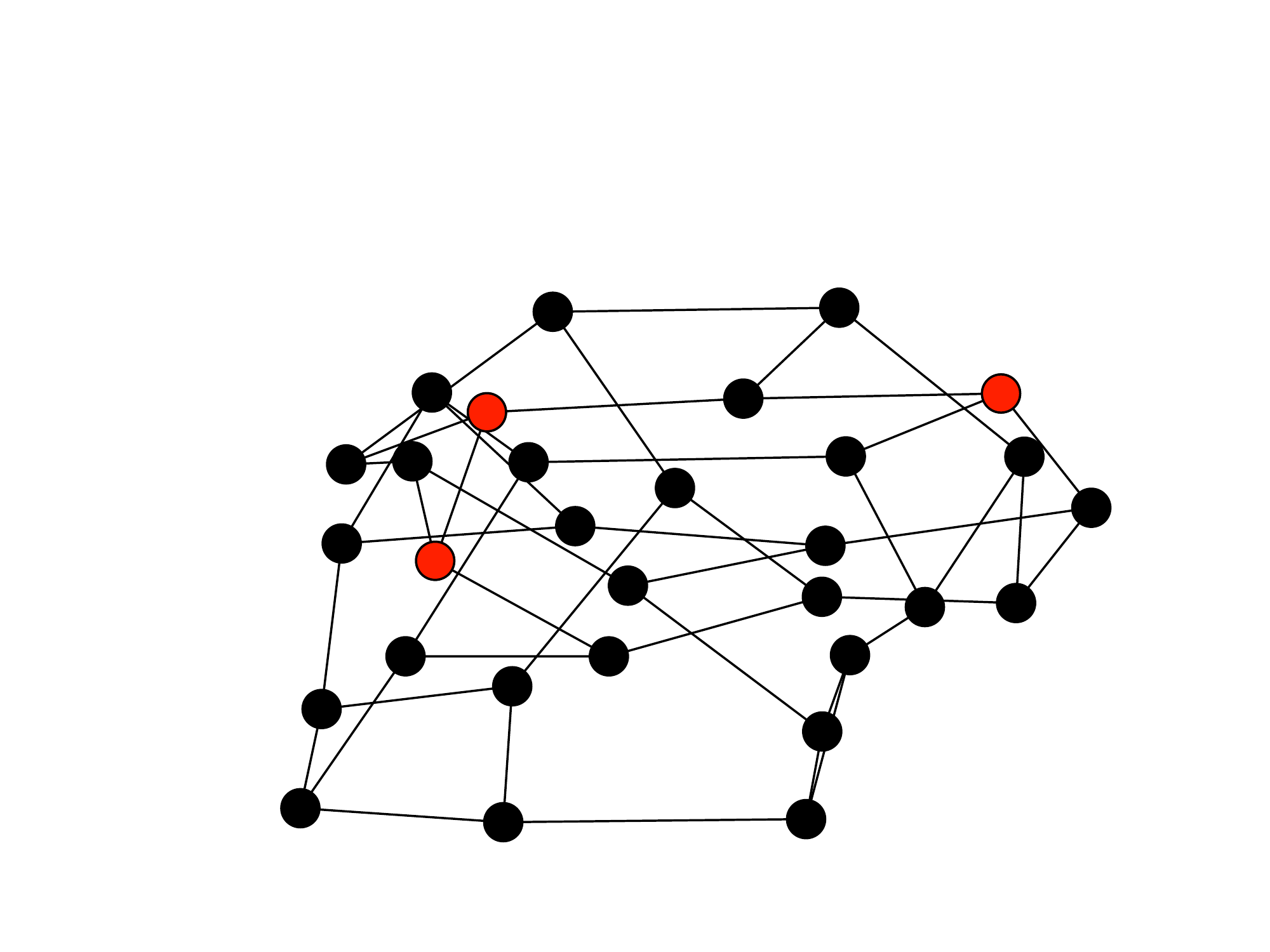}}$\Rightarrow$%
\raisebox{-0.5\height}{\includegraphics[width=0.3\textwidth]{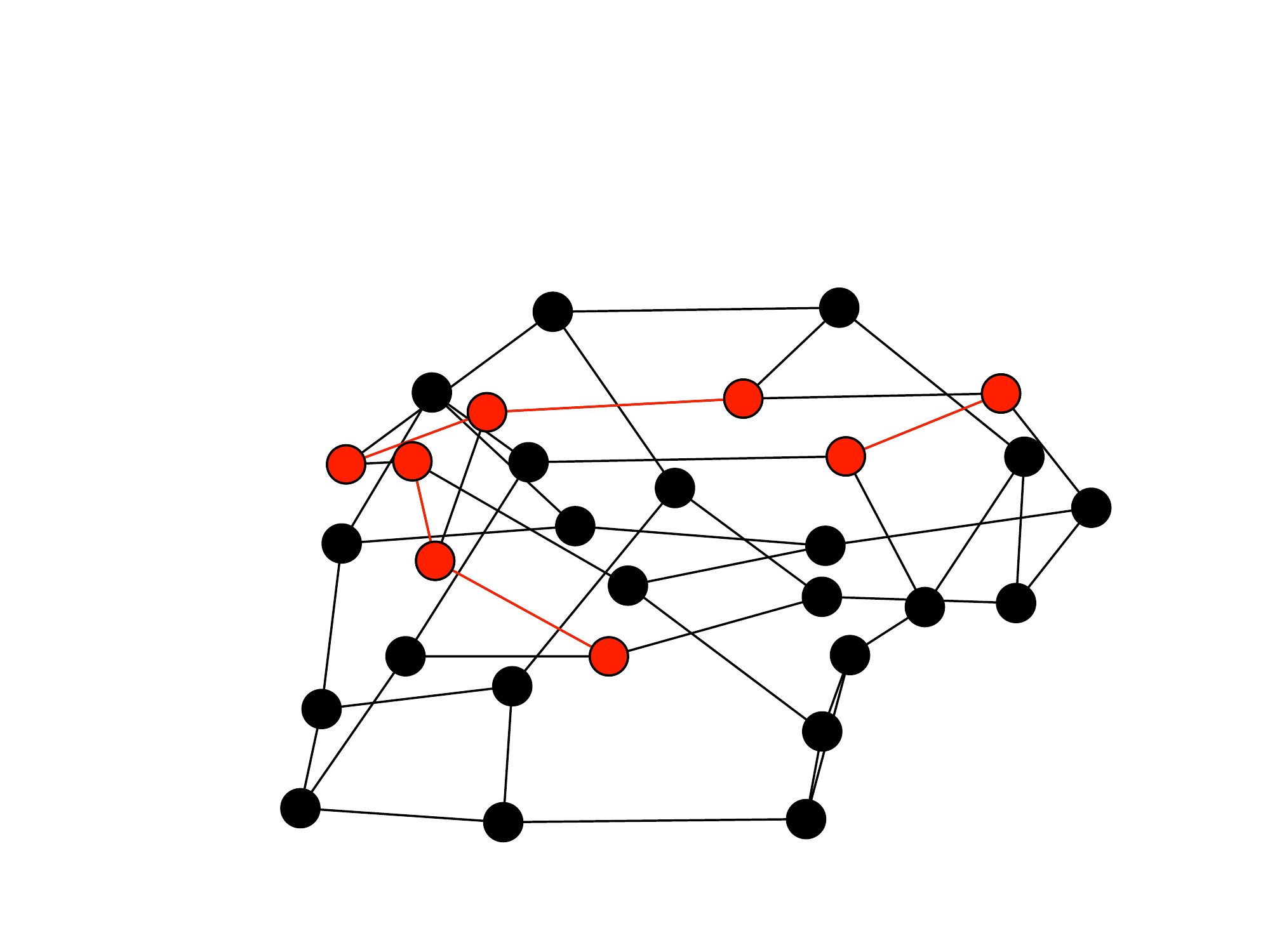}}$\Rightarrow$%
\raisebox{-0.5\height}{\includegraphics[width=0.3\textwidth]{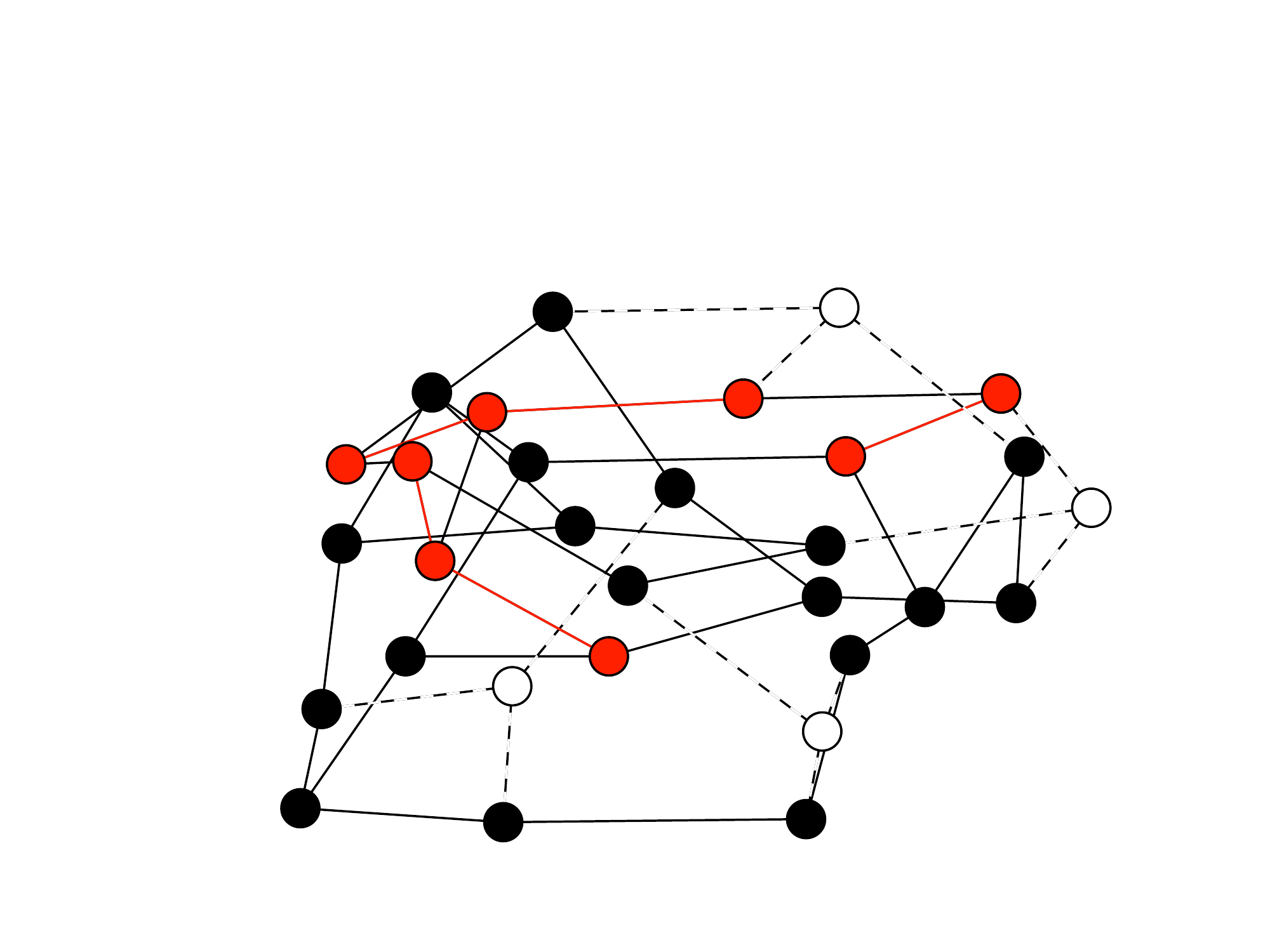}} \\
$\Rightarrow$\raisebox{-0.5\height}{\includegraphics[width=0.3\textwidth]{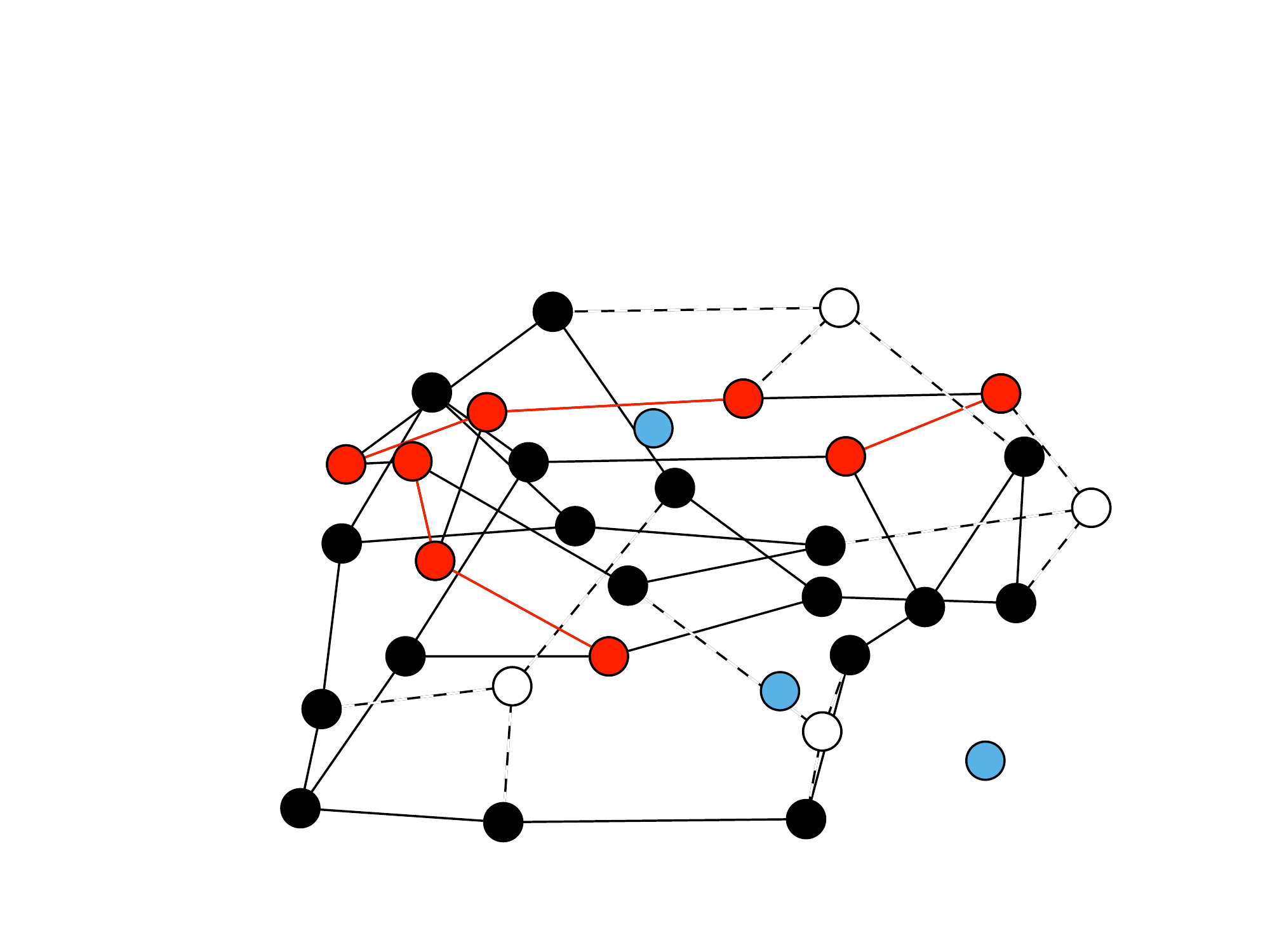}}$\Rightarrow$%
\raisebox{-0.5\height}{\includegraphics[width=0.3\textwidth]{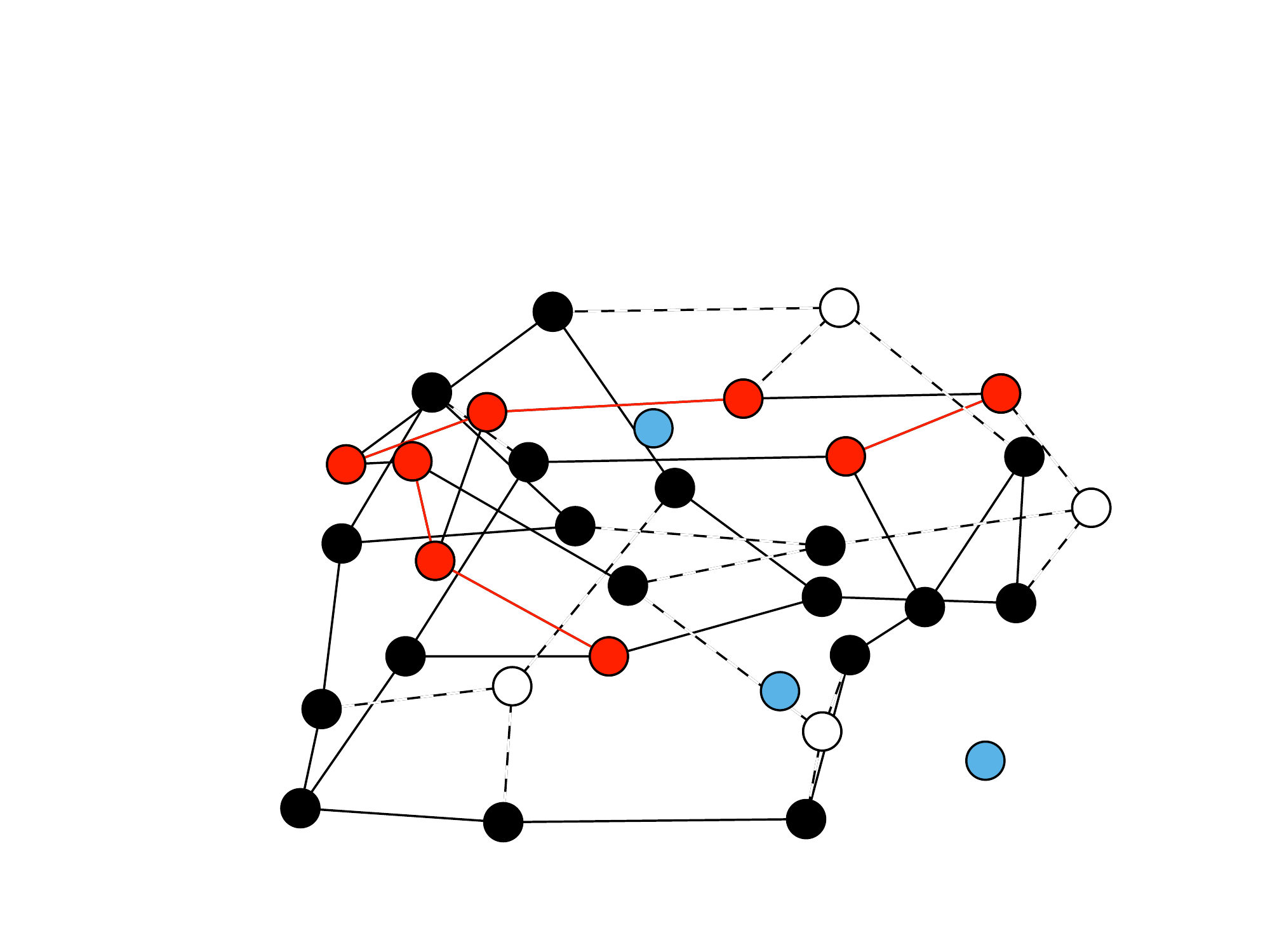}}$\Rightarrow$%
\raisebox{-0.5\height}{\includegraphics[width=0.3\textwidth]{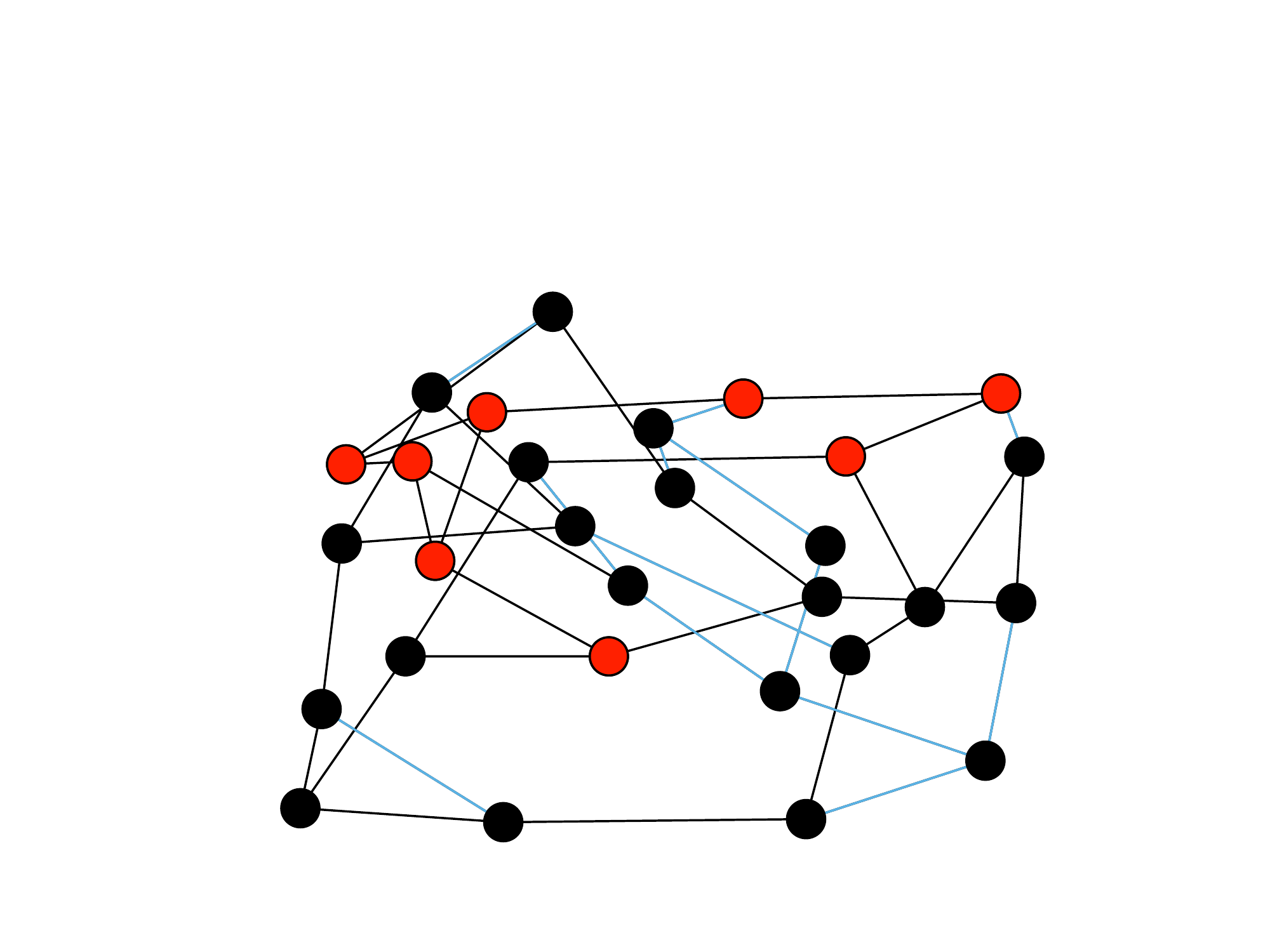}}
\end{center}
\caption{\textbf{Sequence of events in each time step}.  We begin with a network with some infected individuals (red).  Then infected individuals transmit to some partners (red edges).  Then some individuals leave the population (white).  Other individuals are born (blue).  Then some partnerships break (dashed).  Finally new partnerships are created so that the new individuals, the individuals whose partners left, and the individuals whose partnerships broke all return to their target number of partners.  The sequence of events then
  repeats.}
\label{fig:order_of_events}
\end{figure}

We initialize the network as a configuration model network~\cite{newman:structurereview}. 
In each time step actions occur in the following specific order  (illustrated in Fig.~\ref{fig:order_of_events}):
\begin{enumerate}
\item Each partnership connecting a susceptible to an infected individual transmits infection with probability $\tau$.  
\item Individuals leave the population independently with probability $\mu$, freeing their partners to form a replacement partnership in step~\ref{joinedges}.  
\item Next $\mu N$ new individuals arrive where $N$ is the imposed average population size.  Each new individual's ``target degree'', or desired number of partners, $k$ is assigned from the imposed degree distribution with probability $P(k)$.  The individual is given $k$ free ``stubs'' or half-edges which will pair with other stubs to form partnerships. 
\item Each existing partnership ends independently with probability $\eta$, freeing up the two stubs involved. 
\item Free stubs are paired together at random until all individuals reach their target number of partners.  If two stubs are chosen that come from the same individual or would duplicate an existing or a just-terminated partnership, they are left unpaired until the next time step (in the large population limit, this has a negligible impact).\label{joinedges}
\end{enumerate}
This process is then repeated for the next time step.  These assumptions are similar to those of~\cite{leung:demographic,leung:disease}.  Python code implementing these steps is provided as a supplement.  We will present an analytic model that captures the large-population deterministic limit of these assumptions.

For the corresponding simulations, we choose the time step of the discrete-time framework to balance competing interests.  We need a small time step so that $\mu$, $\eta$, and $\tau$ are small (at leading order they are proportional to the time step), otherwise the arbitrary order of events impacts outcomes.  However, for too small of a time step (and also in the continuous-time limit) few partnerships end in a time step.  This makes it difficult for nodes to immediately find new partners.  As the population size increases this becomes less of a problem so smaller time-steps become feasible.  However, the computational effort becomes greater.  

\subsubsection*{Governing equations}

We give an overview of the derivation of the deterministic equations governing the susceptible and infected fractions.  These equations are based on the ``Edge-based Compartmental Modeling'' approach of~\cite{volz:cts_time, miller:volz, miller:ebcm_overview,kiss:EoN}. 

\begin{center}
\begin{table}
\begin{tabular}{|c|c|}
\hline
\textbf{Parameter} & \textbf{Description}\\ \hline\hline
$k$ & \parbox{0.75\textwidth}{\T{}The ``degree'', or number of partners, an individual has (fixed in time).\B{}}\\ \hline
$P(k)$ & \parbox{0.75\textwidth}{\T{}The probability a random individual has degree $k$.\B{}}\\ \hline
$\mu$ & \parbox{0.75\textwidth}{\T{}The probability a random individual will leave the population in a given time step.\B{}}\\ \hline
$\tau$ & \parbox{0.75\textwidth}{\T{}The transmission probability in a time step.\B{}}\\ \hline
$\eta$ & \parbox{0.75\textwidth}{\T{}The probability a partnership will end in a time step.\B{}}\\ \hline
$\rho$ & \parbox{0.75\textwidth}{\T{}The proportion of the population
  randomly infected at $t=0$.\B{}}\\ 
\hline
$b$ & \parbox{0.75\textwidth}{\T{}The number of individuals entering the population each time step (assumed to be constant).\B{}}\\ \hline\hline
$\psi(x)$ & \parbox{0.75\textwidth}{\T{}$\sum_{k=0}^\infty P(k)
  x^k$: The
  probability generating function of the degree distribution.\B{}}
\\ \hline
$p_b$ & \parbox{0.75\textwidth}{\T{}$1-(1-\mu)(1-\eta)$: The probability that a test individual's partnership will end (break) because either the partnership ends naturally (rate $\eta$), or the partner leaves the population (rate $\mu$).  It does not include the possibility of the test individual leaving.\B{}}
\\ \hline
$P_e$ & \parbox{0.75\textwidth}{\T{}$\frac{(1-\mu)(\eta+\mu-\eta\mu)}{(1-\mu)(\eta+\mu-\eta\mu) + \mu} $:
  The probability that a newly
  formed partnership will be with a previously existing individual.\B{}}\\
\hline
$N$ & \parbox{0.75\textwidth}{\T{}$b/\mu$: The average population size.\B{}}\\ \hline
\end{tabular}
\caption{\textbf{The parameters for our simulations and equations.}  The last four are derived from the previous parameters.}
\label{tab:parameters}
\end{table}
\end{center}

\begin{center}
\begin{table}
\begin{tabular}{|c|c|}
  \hline
  \textbf{Variable} & \textbf{Description}\\ \hline \hline 
  $t$ & \parbox{0.75\textwidth}{\T{}Time\B{}}\\ \hline
  $u$ & \parbox{0.75\textwidth}{\T{}The test individual\B{}}\\ \hline
  $a_u$ & \parbox{0.75\textwidth}{\T{}The age of test individual $u$,
    measured so that $a_u=0$ in the first time step after $u$ is born.\B{}}\\
  \hline
  $a_e$ & \parbox{0.75\textwidth}{\T{}Age of an partnership of interest,
    measured so that $a_e=0$ in the first time step after the partnerships
    forms.\B{}}\\ \hline
   $S(t)$ & \parbox{0.75\textwidth}{\T{}The proportion of the
     population that is susceptible at time $t$, equivalently the
     probability a randomly selected individual is susceptible at time
     $t$, or equivalently the probability a test individual is
     susceptible at time $t$.\B{}}\\ \hline
  $I(t)$ & \parbox{0.75\textwidth}{\T{}$1-S(t)$: The proportion
    infected at time $t$.\B{}}\\ \hline
  $s(t,a_u)$ & \parbox{0.75\textwidth}{\T{}The probability a test
    individual of age $a_u$ is susceptible at time $t$.\B{}}\\ \hline
  $\Theta(t,a_u)$ & \parbox{0.75\textwidth}{\T{}The probability a stub
    belonging to $u$ has not transmitted infection to it from a partner
    by the start of time step $t$.\B{}}  \\ \hline
  $\Phi_{S}(t,a_u)$ & \parbox{0.75\textwidth}{\T{}As for $\Theta$ (no partner has transmitted to $u$ through the stub), but with the additional requirement that at the start of time step $t$ the partner is susceptible.\B{}}\\ \hline
  $\Phi_{I}(t,a_u)$ & \parbox{0.75\textwidth}{\T{}As for $\Theta$ (no partner has transmitted to $u$ through the stub), but
    with the additional requirement that at the start of time step $t$
    the partner is infected.\B{}}\\ \hline
  $\phi_{S}(t,a_u,a_e)$ &  \parbox{0.75\textwidth}{\T{}The probability
    that a stub belonging to an age $a_u$ individual has not transmitted infection to it by time $t$, is connected to a susceptible partner, and the current partnership (or ``edge'') has age $a_e$.\B{}}\\ \hline
$\chi(t,a_e)$ & \parbox{0.75\textwidth}{\T{}The probability that
  an age $a_e$ partnership (or ``edge'') of a test individual connects to a susceptible
  individual.\B{}}\\ \hline
\end{tabular}
\caption{\textbf{The variables for our equations.}}
\label{tab:variables}
\end{table}
\end{center}

Tables~\ref{tab:parameters} and~\ref{tab:variables} summarize the
parameters and variables of the model respectively.  The model parameters are the initial fraction infected $\rho$, the per-time-step death probability $\mu$, number of births $b$, transmission probability $\tau$, and partnership change probability $\eta$.  In addition to time $t$, the key independent variables are the age of an individual and the age of a partnership.  The age of an individual gives some information about that individual's status (a recent entrant is generally less likely to be infected than one who has been in the population for a while).  In addition the age of a partnership gives information about the age of the partner (a partnership cannot be ``older'' than the partner).  In calculating the risk an individual has from its partners, we need to account for the probability the partner has a given status.  This depends on the age of the partner, which in turn depends on the age of the partnership, which itself is dependent on the age of the individual.  To sort out the dependencies, the individual age and the partnership age are needed as independent variables.  The resulting equations are low-dimensional, but are significantly more involved than even the dynamic network models presented in~\cite{miller:ebcm_overview}.  

We seek the susceptible and infected fractions of the population $S$ and $I$.  We outline a simplified derivation ignoring some details of the initial condition and assuming the population has equilibrium size $N = b/\mu$.  The probability that a random individual $u$ has age $a_u$ is $\mu(1-\mu)^{a_u}$ because the proportion born in any time step is $\mu$ (it must balance the proportion that die) and the probability of surviving $a_u$ time steps is $(1-\mu)^{a_u}$.  The probability that a random individual of age $a_u$ and with $k_u$ partners is susceptible is $\Theta(t,a_u)^{k_u}$ where $\Theta(t,a_u)$ is the probability a random partner (or its predecessors along the same stub) has not transmitted to $u$.  Thus the probability a random individual is susceptible is
\[
S(t) = \mu\sum_{a_u=0}^\infty \left((1-\mu)^{a_u} \sum_k P(k) \Theta(t, a_u)^k\right)
\]
Introducing the probability generating function $\psi(x) = \sum_k P(k) x^k$, gives $s(t,a_u) = \psi(\Theta(t,a_u))$ is the probability an age $a_u$ individual is susceptible.  This expression then simplifies to $S(t) = \mu \sum (1-\mu)^{a_u} s(t,a_u)$.  The probability a random individual is infected is $I(t) = 1-S(t)$.  

We now derive an equation for $\Theta$ by noting that
$\Theta(t,a_u) = \Theta(t-1, a_u-1) - \tau \Phi_I(t-1, a_u-1)$ where
$\Phi_I$ is the probability that the partner is infected and neither it nor any predecessor has transmitted to $u$.  We find $\Phi_I$ by first solving for $\Phi_S$, the probability that the partner (and predecessors) have not transmitted and the partner is susceptible, and using $\Phi_I = \Theta-\Phi_S$.  The derivation for $\Phi_S$ is given in the SI.  It is similar to the derivation of $S$, but requires additional handling of the possible ages of the partner because the age distribution of the partners is different for partnerships of different ages.

Once we incorporate the initial condition and the full details of deriving $\Phi_S$, the governing equations are
\begin{align*}
S(t) &= \mu \sum_{a_u=0}^{\infty} (1-\mu)^{a_u} s(t,a_u)\\
s(t,a_u) &= \begin{cases} \psi(\Theta(t,a_u)) & a_u < t\\
     (1-\rho)\psi(\Theta(t,t)) & a_u \geq t
     \end{cases}\\
I(t) &= 1-S(t)\\
\Theta(t,0)&=1\\
\Theta(0,a_u)&=1\\
\Theta(t,a_u) &= \Theta(t-1,a_u-1) - \tau \Phi_I(t-1,a_u-1) \qquad t, a_u\geq 1\\
\Phi_I(t,a_u) &= \Theta(t,a_u)-\Phi_S(t,a_u) \\
\Phi_S(t,a_u) &= (1-p_b)^{a_u} \phi_S(t,a_u,a_u) + p_b \sum_{a_e=0}^{a_u-1} (1-p_b)^{a_e}\phi_S(t,a_u,a_e)\\
\phi_S(t,a_u,a_e) &= \Theta(t-a_e, a_u-a_e) \chi(t,a_e)\\
\chi(t,a_e) &= \begin{cases} (1-\rho)
  \frac{\psi'(\Theta(t,t))}{\ave{K}} & a_e \geq t\\[12pt]
(1-P_e) \frac{\psi'(\Theta(t,a_e))}{\ave{K}}\\
\quad+ P_e \mu {\displaystyle\sum_{a_v=a_e+1}^{t-1}} (1-\mu)^{a_v-a_e-1}\Theta(t-a_e,a_v-a_e)\frac{\psi'(\Theta(t,a_v))}{\ave{K}} & a_e < t\\
\quad+ P_e (1-\rho)\Theta(t-a_e,t-a_e)\frac{\psi'(\Theta(t,t))}{\ave{K}} (1-\mu)^{t-a_e-1}
\end{cases}
\end{align*}
We provide a full derivation of this model and give a continuous-time differential equations version in the Supporting Information.
\subsubsection*{Specific modeled population}
Although the equations allow different individuals to have a different
number of partners, except where specifically noted we focus on populations in which all individuals have the same number of partners.  In particular, this eliminates the need to consider how per-partnership transmission rates may depend on the individual's number of partners.  This allows us to focus on concurrency without the effect of some individuals having more frequent sexual activity than others.

So for our comparisons, $\psi(x) = x^k$, \ $\ave{K} = k$ for some fixed value $k$.  As a base case, we consider serial monogamy where each individual has a single partner ($k=1$).  Transmission occurs in a time-step with probability $\tau_1$ and partnerships end with probability $\eta_1$.  Individuals leave the population with probability $\mu$.  

We compare this with homogeneous populations having concurrency.  We assume that for different values of $k$ the  populations are arranged such that the cumulative number of partners an individual has over a long period of time is the same.  So populations with smaller $k$ must have faster turnover.  This implies that $\eta = \eta_1/k$ so that the $k$ partnerships each lasts $k$ times as long.  Similarly we assume that the expected number of transmissions an infected individual would cause in a time step is the same.  This requires $\tau = \tau_1/k$.  

\subsubsection*{Advantages of an analytic model}
There are a number of important benefits to having an analytic model (even if it can only be solved numerically) as opposed to relying on stochastic simulation.  In general, an analytic model allows us to gain much more insight into a system because the mathematical relationships that emerge can often be interpreted as an interaction between different physical effects from which we understand how system behavior emerges.  In contrast, with stochastic simulation it is difficult to extract those mathematical relationships.

We highlight several practical advantages of analytic models.
\begin{itemize}
\item If our goal is to predict the large-population behavior, a single stochastic simulation may take much longer and use much more memory than a numerical solution of the analytic equations: even in the simplest well-mixed homogeneous population, we must have enough individuals in the simulation to accurately capture the average, while the numerical solution only needs to track the average.  In our comparisons the numerical solution can run hundreds of times faster than a large stochastic simulation, and use several orders of magnitude less memory. 
\item If we want to see how the equilibrium size changes as parameters change, we can simulate to equilibrium, then change the parameters a bit and simulate further watching the system relax to the new equilibrium.  With the numerical equations, we can do something similar, but once we identify the equilibria for multiple parameter values, we can then give good estimates of the equilibrium variables at new parameters.  This can be fed into the numerical solution as an initial condition, allowing for dramatically faster convergence.  We cannot use the same sort of extrapolation to find good initial conditions for a new stochastic simulation.
\item The analytic model allows for the application of mathematical
  tools that are difficult to adapt to simulations.  For example, the numerical solution of the differential equations version can use adaptive step size and other numerical techniques.  We can look for properties of equilibria by assuming that the analytic variables do not change. We can look for relationships between the parameters which determine when epidemic growth is possible.
\end{itemize}

\section*{Results}

In this section we briefly compare the analytic model with simulations to show that the analytic model gives accurate predictions.  We then use the analytic model to investigate the impact of concurrency on the endemic equilibrium and on the early growth rates.  The impact of concurrency saturates in both cases, with the impact on the endemic equilibrium saturating at lower levels of concurrency than the impact on early growth.

\subsubsection*{Accuracy of the analytic model}

\begin{figure}
\includegraphics[width=0.48\textwidth]{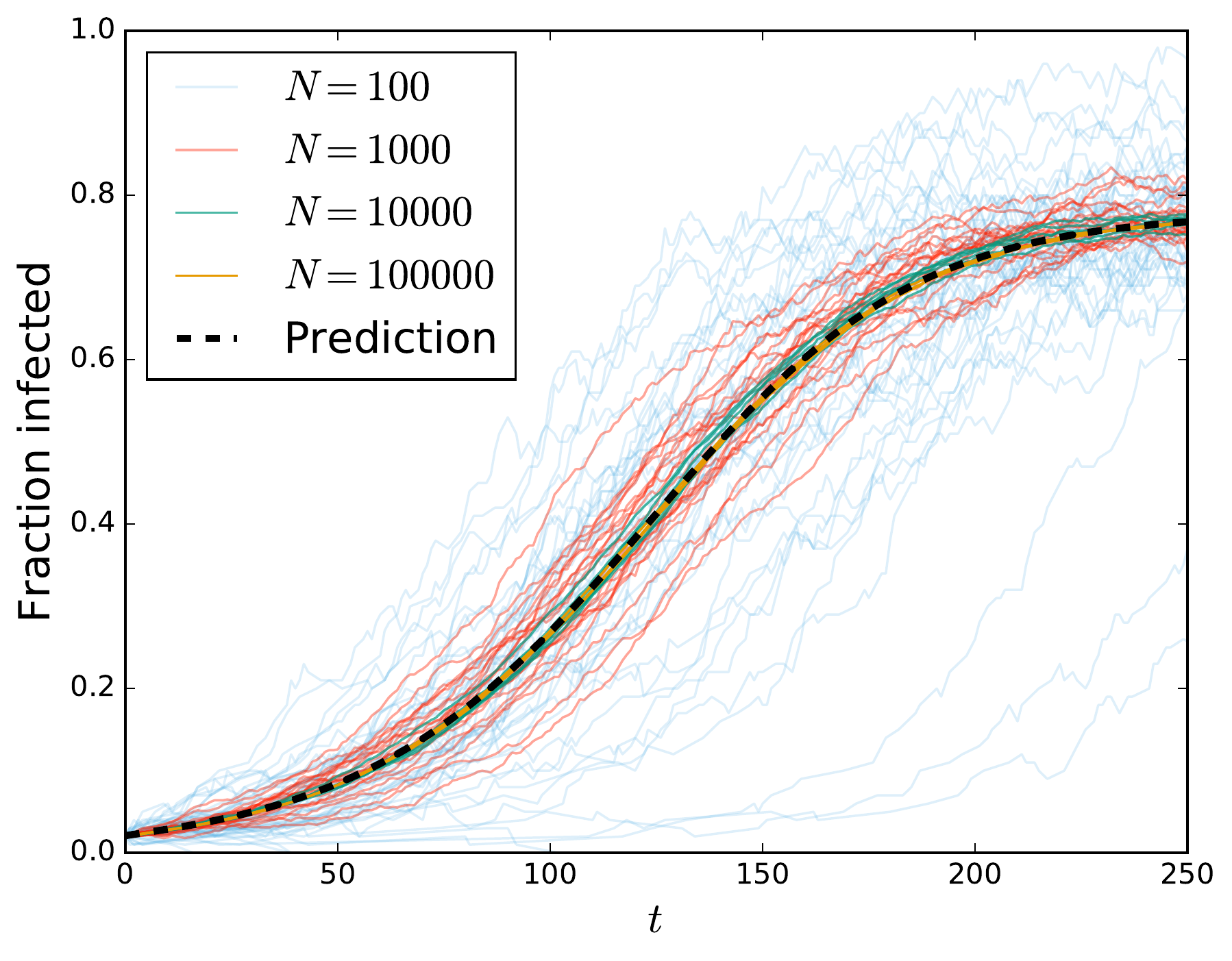} \hfill
\includegraphics[width=0.48\textwidth]{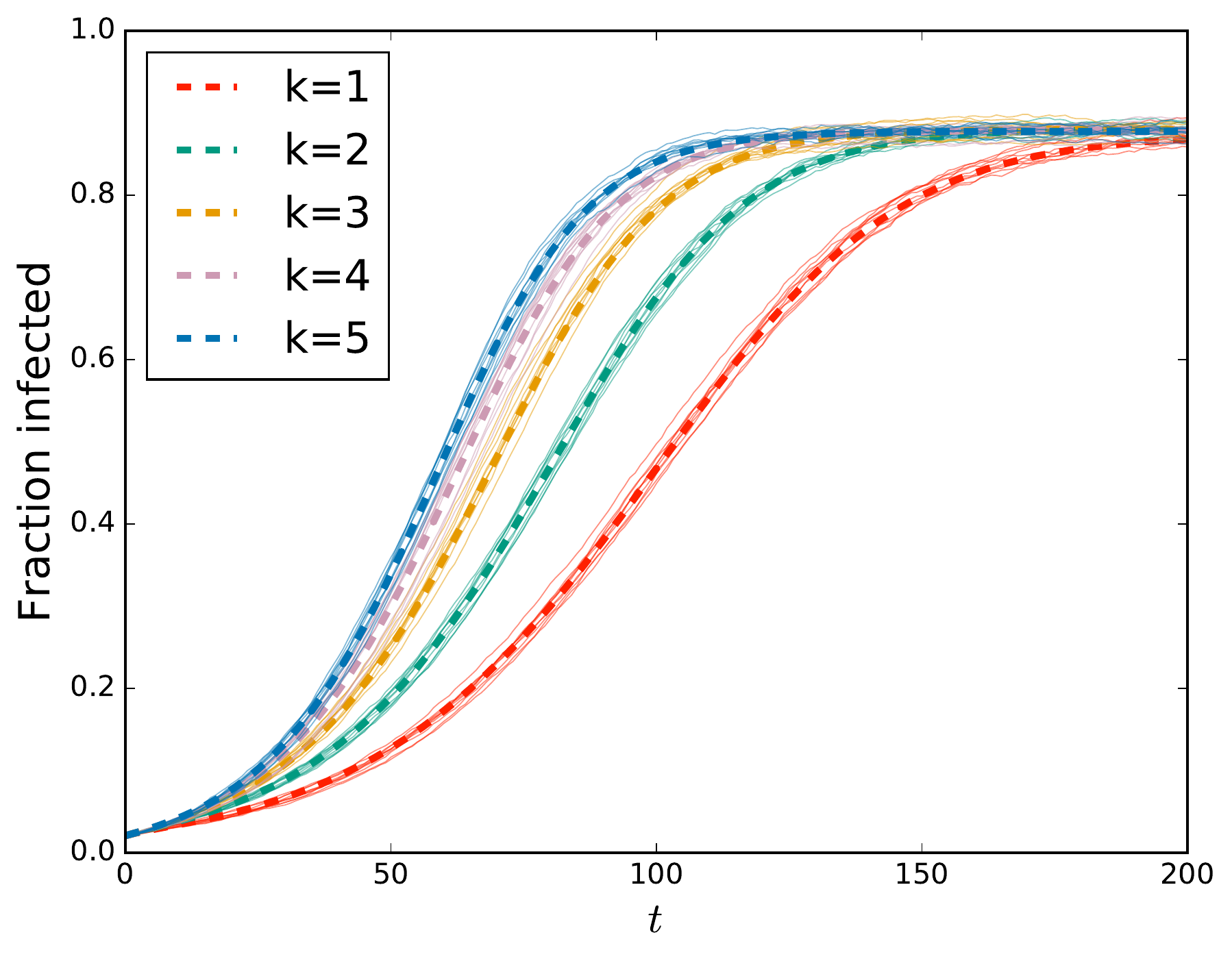}
\caption{\textbf{Comparison of analytic predictions and
    stochastic simulations:}  (Left) Stochastic epidemic
  simulations in populations of average size $N = 10^2$, $10^3$, $10^4$
  and $10^5$ with $k=3$, \ $\eta=0.2/3$, \ $\tau = 0.05/3$, \
  $\rho=0.02$, and $\mu=0.01$.  As $N$ increases, the simulations
  converge to the analytic prediction.  (Right) Comparison of analytic predictions and simulations for different
  values of $k$, with $\eta_1=0.1$, \
  $\tau_1=0.1$, \ $\rho=0.02$, \ $\mu=0.01$ and average population
  size $N=10^4$ with $\tau=\tau_1/k$ and $\eta = \eta_1/k$.  We find
  excellent agreement between predictions (thick dashed curves) and stochastic simulations (thin solid curves).}
\label{fig:theory_vs_pred}
\end{figure}

\begin{figure}
\includegraphics[width=0.48\textwidth]{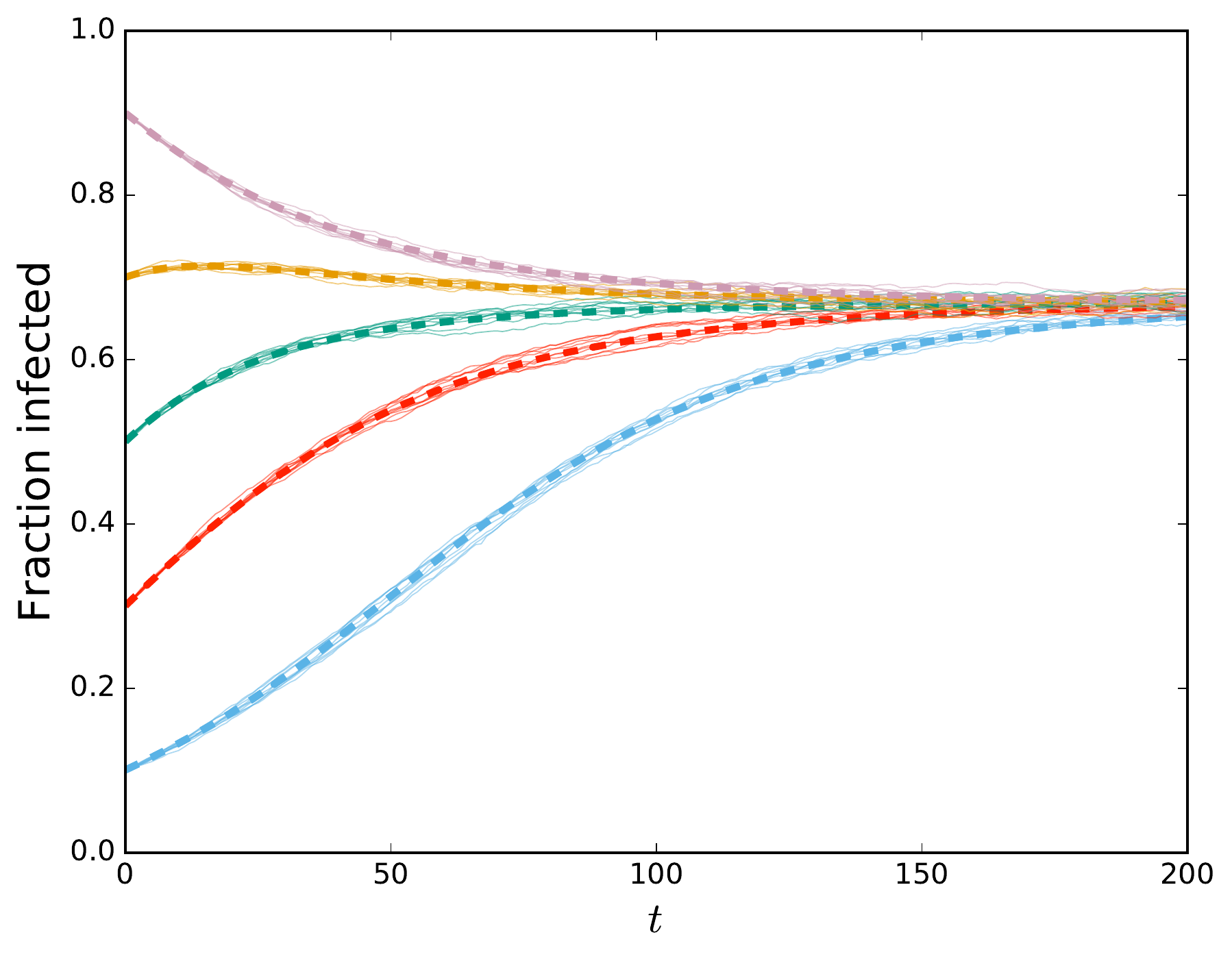} \hfill 
\includegraphics[width=0.48\textwidth]{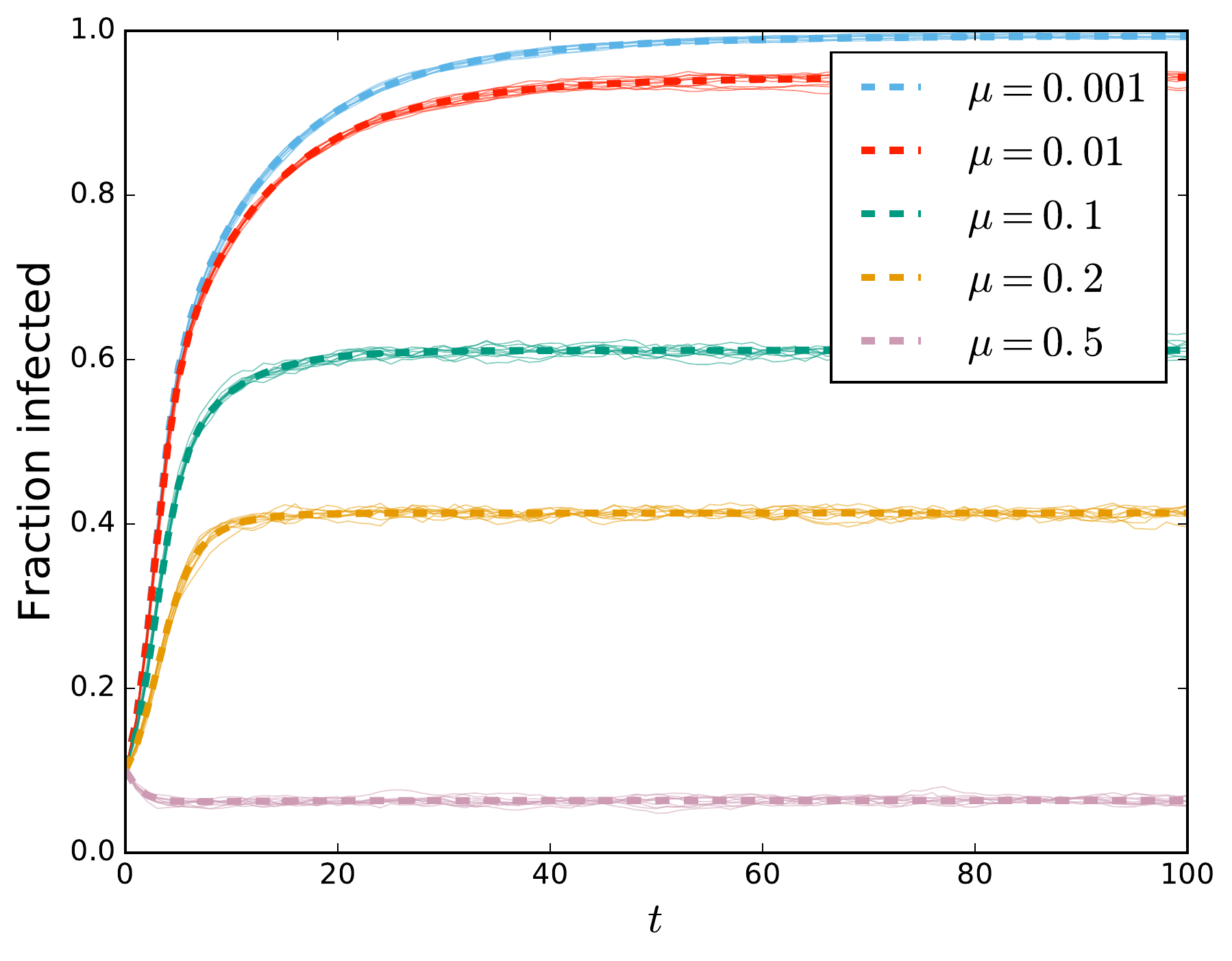}
\caption{\textbf{Comparison of dynamics from analytic predictions and stochastic simulations in populations with heterogeneous degree:}  (Left) Disease spread in populations with average size $N = 10^4$ and degree probabilities $P(2)=P(7)=1/2$.  The parameters are $\tau = 0.01$, \ $\eta = 0.005$, and $\mu = 0.01$.  The initial fraction infected $\rho$ varies.  (Right) Disease spread in populations with average size $N=10^4$ and degree probabilities $P(1)=1/2$, \ $P(10)=1/3$, and $P(20)=1/6$.  The parameter $\mu$ varies between populations.  The remaining parameters are $\eta = 0.05$ and $\tau = 0.1$.  In both plots the dashed curves are analytic predictions and thin solid curves are stochastic simulations.}
\label{fig:heterogeneous}
\end{figure}

Figs.~\ref{fig:theory_vs_pred} and~\ref{fig:heterogeneous} compare solutions of the analytic model with stochastically simulated epidemics across a range of conditions.  
The left plot of Fig.~\ref{fig:theory_vs_pred} shows that as the population size increases, stochastic simulations converge to the predicted dynamics.  The right plot considers populations in which all individuals have $k$ partners for $k=1$, $2$, $3$, $4$, and $5$ with $\tau=\tau_1/k$ and $\eta=\eta_1/k$ for $\eta_1=\tau_1=0.1$ and $\mu = 0.01$.  Simulations and predictions are a good match for different values of $k$. 

Figure~\ref{fig:heterogeneous} looks at disease
spread in populations with heterogeneous degrees, again showing excellent agreement between stochastic simulations and analytic predictions. 

Our main conclusion from Figs.~\ref{fig:theory_vs_pred} and~\ref{fig:heterogeneous} is that the equations accurately predict the large-$N$ dynamics of simulations regardless of the parameters used.  

\subsubsection*{Further observations}

From the left plot in Fig.~\ref{fig:heterogeneous} we infer that for a given population, the initial proportion infected $\rho$ does not influence the final state.  At small $\rho$, the early dynamics are dominated by new infections to high- and low-degree nodes.  At large $\rho$ they are dominated by removal of infected high- and low-degree nodes.  Interestingly, we see that for some intermediate values of $\rho$ the prevalence initially grows and then decays.   At these intermediate values, initially more high-degree nodes are being infected than are leaving while more low-degree nodes are leaving than being infected.  Thus the two sets of nodes have opposing impacts on the dynamics.  In this case, the growth in high-degree infections is initially the dominant effect, but it saturates while many low-degree infected individuals are still being removed.

From the right plot of Fig.~\ref{fig:heterogeneous}, we infer that increasing the population turnover rate decreases the proportion infected.  This is not particularly surprising as it implies that infected individuals leave the population sooner, having had less opportunity to cause further infections.

\subsubsection*{Impact of concurrency}
\begin{figure}
\includegraphics[width=\textwidth]{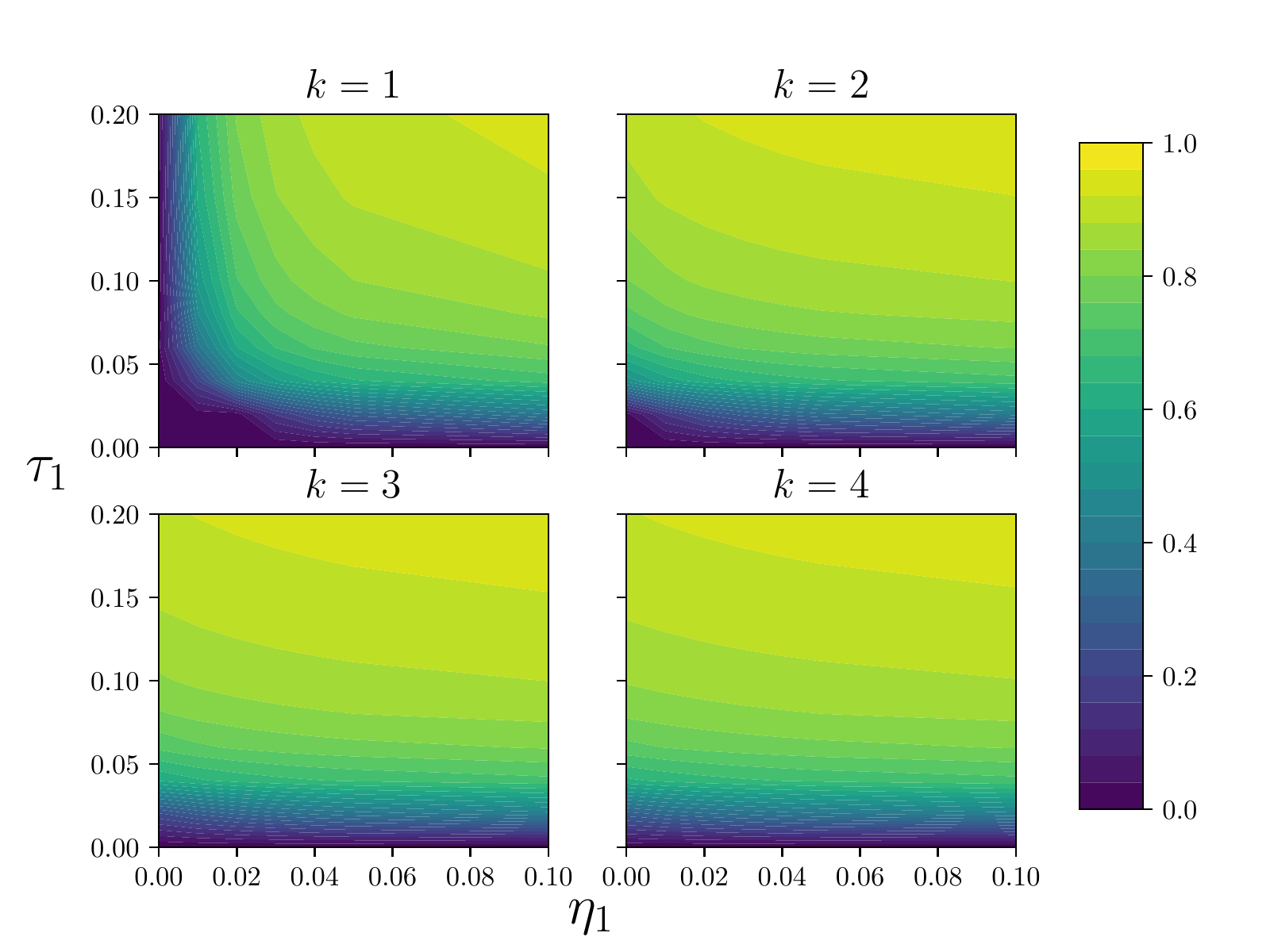}

\caption{\textbf{Comparison of equilibrium sizes for different values of $k$:}  The equilibrium fraction infected for $\mu=0.01$ and different $\eta_1$ and $\tau_1$.  We use the same axes $\eta_1$ and $\tau_1$, taking $\eta=\eta_1/k$ and $\tau = \tau_1/k$.  (Top left) $k=1$, (Top right) $k=2$, (Bottom left) $k=3$, and (Bottom right) $k=4$.  As $k$ increases, the figures converge: the effect of concurrency on the equilibrium size quickly saturates.}
\label{fig:contours}
\end{figure}

\begin{figure}
\includegraphics[width=\textwidth]{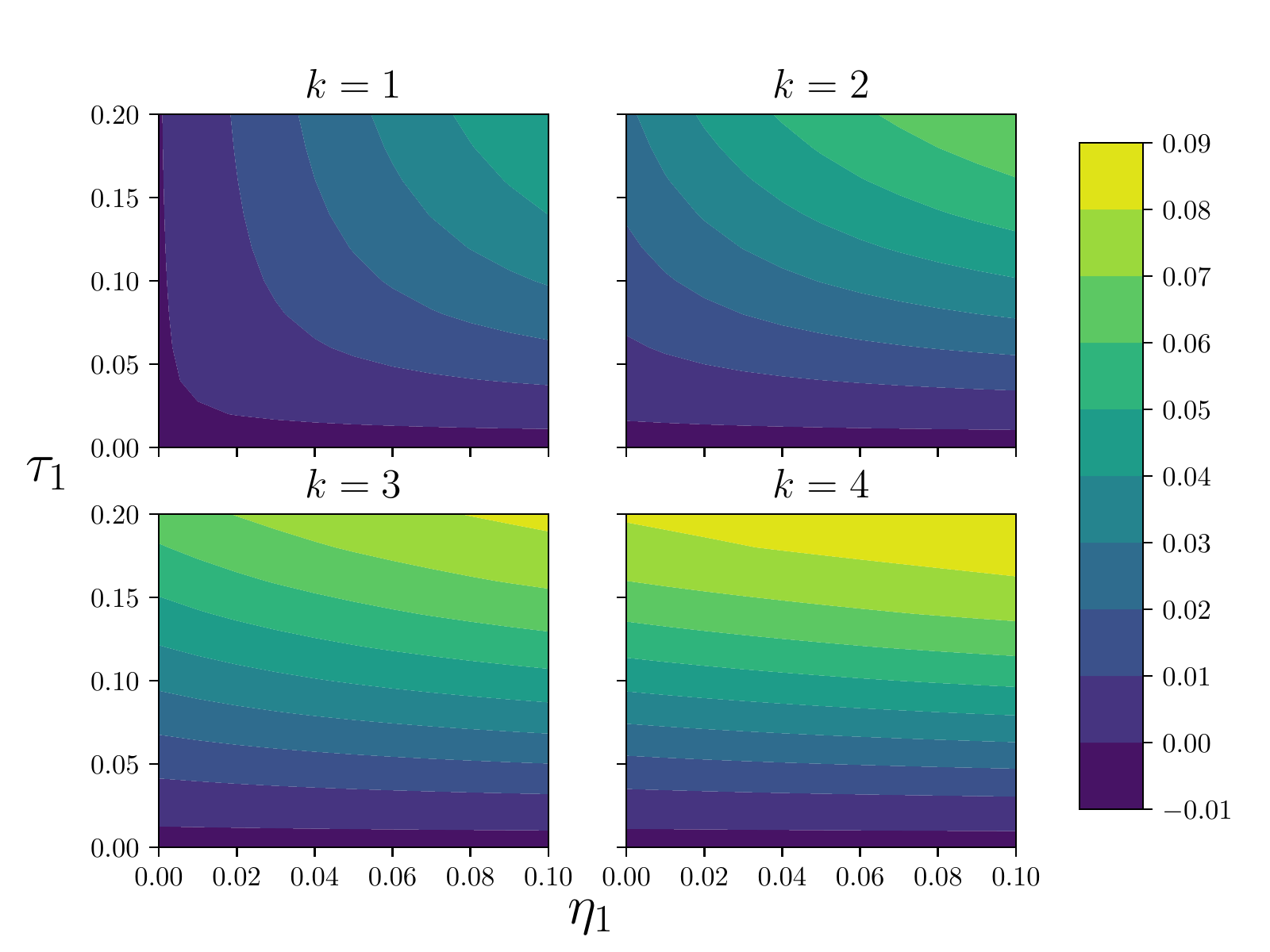}
\caption{\textbf{Comparison of early growth with different levels of concurrency:}  Contour plots of $\Delta I/I$ once the dynamics have entered the early exponential phase.  (Top left) $k=1$, (Top right) $k=2$, (Bottom left) $k=3$, and (Bottom right) $k=4$.  Although eventually the change saturates with larger $k$, it does not saturate as quickly as the equilibrium size does.}
\label{fig:early_growth}
\end{figure}

We now specifically explore the model predictions as we change the amount of concurrency.  The right-hand plot of Fig.~\ref{fig:theory_vs_pred} provides a special case of our more
generic results.  
Interestingly we see that the equilibrium level does not vary significantly as concurrency increases, but the early growth rate does.  
Figs.~\ref{fig:contours} and~\ref{fig:early_growth} show how the
equilibrium infection levels and early growth rates change as $k$
changes for a range of values of $\eta_1$ and $\tau_1$.

Fig.~\ref{fig:contours} 
shows that the impact of concurrency on the equilibrium epidemic size can be significant, but that the effect of increasing $k$ may saturate quickly.  The impact of concurrency is greatest if the partnership duration is long ($\eta_1$ small).  For much of parameter space, Fig.~\ref{fig:contours} suggests that once a little concurrency is present, increased concurrency has little further effect on the equilibrium size.

In Fig.~\ref{fig:early_growth}, we demonstrate that in our model concurrency has a larger effect on the growth of the epidemic than on the equilibrium size.  Although the impact of increasing concurrency on the early growth eventually saturates, it does so at larger $k$ than for equilibrium size.

Figures~\ref{fig:contours} and~\ref{fig:early_growth} both show that saturation occurs soonest at lower transmission rates and higher partnership turnover rates, where the probability of multiple transmissions in a partnership is small.

\section*{Discussion}

We now discuss mechanisms whereby the impact of concurrency saturates, explore the implications for intervention design, and 
list caveats because of important effects neglected in our model.

We designed our comparisons so that, 
for different $k$, the number of partners an individual has over a long period of time and the total number of interactions within each partnership are the same.  Thus we know that the effects we observe are not explained by within-population heterogeneity in degree, within-population heterogeneity in sexual activity rates, between-population differences in typical life-time number of partners, or between-population differences in the number of transmissions an individual causes per time step.  All of these effects have been removed.  Each population is homogeneous and they differ only in the number of concurrent partnerships.

To understand why concurrency does or does not matter in the different
cases we take the perspective of the disease, observing transmission
events and their outcomes.  If concurrency has a significant impact on
the population-scale spread of disease, it must be possible to infer
the existence of concurrency by exploring the population structure in
the same way the disease spreads.  So we ask ourselves, ``how easily
can the disease measure the concurrency?''

\subsubsection*{Low transmission rates/fast partnership turnover}
In our model, we observed that concurrency has a reduced effect at
lower transmission rate or if the partnership turnover rate is large.
We now investigate why this is which will help us identify whether
these observations should hold in the real world.    

Looking back at Fig.~\ref{fig:concurrency_display}, we see that if we ignore the shading that designates when a partnership is in existence or not, it is still possible to infer which cases correspond to concurrent relationships by looking at the dashed lines that represent potential transmission events (that is interactions that would cause infection if the recipient were susceptible).   The most obvious sign of concurrency is that at least one transmission event in one partnership lies between sequential transmission events in the other partnership.

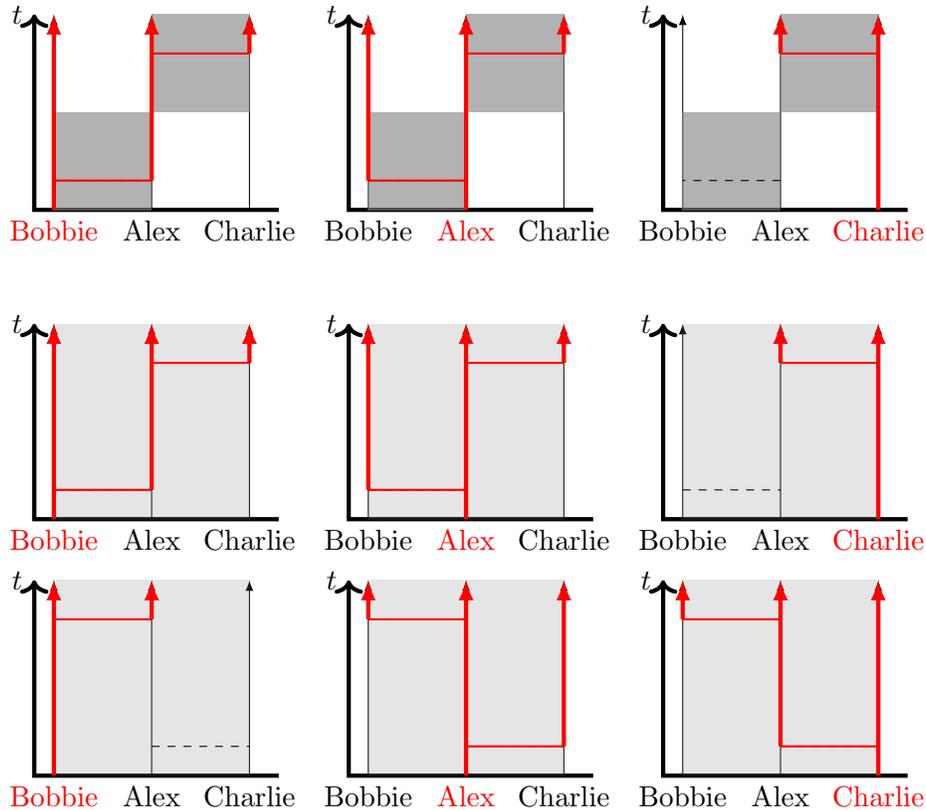
\begin{figure}
\begin{center}
\begin{tikzpicture}[scale=1.3] 
\axes;
\person{\xB}{{\color{red}Bobbie}};
\person{\xA}{Alex};
\person{\xC}{Charlie};

\partnership{\xB}{\xA}{0}{0.5*\ymax}{0.6};
\partnership{\xA}{\xC}{1}{\ymax}{0.6};

\foreach \y in {0.3}   
   {\FailTrans{\xB}{\xA}{\y};};
\foreach \y in {1.6}
   {\FailTrans{\xA}{\xC}{\y};};

\InfectionStart{\xB}{0};
\SuccessTrans{\xB}{\xA}{0.3};
\InfectionStart{\xA}{0.3};
\SuccessTrans{\xA}{\xC}{1.6};
\InfectionStart{\xC}{1.6}; 
\end{tikzpicture}
\begin{tikzpicture}[scale=1.3]
\axes;
\person{\xB}{Bobbie};
\person{\xA}{{\color{red}Alex}};
\person{\xC}{Charlie};

\partnership{\xB}{\xA}{0}{0.5*\ymax}{0.6};
\partnership{\xA}{\xC}{1}{\ymax}{0.6};

\foreach \y in {0.3}
   {\FailTrans{\xA}{\xB}{\y};};
\foreach \y in {1.6}
   {\FailTrans{\xC}{\xA}{\y};};

\InfectionStart{\xA}{0};
\SuccessTrans{\xC}{\xA}{1.6}; 
\InfectionStart{\xC}{1.6};
\SuccessTrans{\xB}{\xA}{0.3}; 
\InfectionStart{\xB}{0.3};
\end{tikzpicture}
\begin{tikzpicture}[scale=1.3]
\axes;
\person{\xB}{Bobbie};
\person{\xA}{Alex};
\person{\xC}{{\color{red}Charlie}};

\partnership{\xB}{\xA}{0}{0.5*\ymax}{0.6};
\partnership{\xA}{\xC}{1}{\ymax}{0.6};

\foreach \y in {0.3}
   {\FailTrans{\xA}{\xB}{\y};};
\foreach \y in {1.6}
   {\FailTrans{\xC}{\xA}{\y};};

\InfectionStart{\xC}{0};
\SuccessTrans{\xC}{\xA}{1.6}; 
\InfectionStart{\xA}{1.6};
\end{tikzpicture}\\[20pt]

\begin{tikzpicture}[scale=1.3]
\axes;
\person{\xB}{{\color{red}Bobbie}};
\person{\xA}{Alex};
\person{\xC}{Charlie};

\partnership{\xB}{\xA}{0}{\ymax}{0.2};
\partnership{\xA}{\xC}{0}{\ymax}{0.2};

\foreach \y in {0.3}   
   {\FailTrans{\xB}{\xA}{\y};};
\foreach \y in {1.6}
   {\FailTrans{\xA}{\xC}{\y};};

\InfectionStart{\xB}{0};
\SuccessTrans{\xB}{\xA}{0.3}; 
\InfectionStart{\xA}{0.3};
\SuccessTrans{\xA}{\xC}{1.6};
\InfectionStart{\xC}{1.6}; 

\end{tikzpicture}
\begin{tikzpicture}[scale=1.3]
\axes;
\person{\xB}{Bobbie};
\person{\xA}{{\color{red}Alex}};
\person{\xC}{Charlie};

\partnership{\xB}{\xA}{0}{\ymax}{0.2};
\partnership{\xA}{\xC}{0}{\ymax}{0.2};

\foreach \y in {0.3}   
   {\FailTrans{\xB}{\xA}{\y};};
\foreach \y in {1.6}
   {\FailTrans{\xA}{\xC}{\y};};
\InfectionStart{\xA}{0};
\SuccessTrans{\xC}{\xA}{1.6};
\InfectionStart{\xC}{1.6};
\SuccessTrans{\xA}{\xB}{0.3};
\InfectionStart{\xB}{0.3}; 
\end{tikzpicture}
\begin{tikzpicture}[scale=1.3]
\axes;
\person{\xB}{Bobbie};
\person{\xA}{Alex};
\person{\xC}{{\color{red}Charlie}};

\partnership{\xB}{\xA}{0}{\ymax}{0.2};
\partnership{\xA}{\xC}{0}{\ymax}{0.2};

\foreach \y in {0.3}   
   {\FailTrans{\xB}{\xA}{\y};};
\foreach \y in {1.6}
   {\FailTrans{\xA}{\xC}{\y};};
\InfectionStart{\xC}{0};
\SuccessTrans{\xC}{\xA}{1.6};
\InfectionStart{\xA}{1.6};

\end{tikzpicture}\\

\begin{tikzpicture}[scale=1.3]
\axes;
\person{\xB}{{\color{red}Bobbie}};
\person{\xA}{Alex};
\person{\xC}{Charlie};

\partnership{\xB}{\xA}{0}{\ymax}{0.2};
\partnership{\xA}{\xC}{0}{\ymax}{0.2};

\foreach \y in {1.6}   
   {\FailTrans{\xB}{\xA}{\y};};
\foreach \y in {0.3}
   {\FailTrans{\xA}{\xC}{\y};};

\InfectionStart{\xB}{0};
\SuccessTrans{\xB}{\xA}{1.6}; 
\InfectionStart{\xA}{1.6};
\end{tikzpicture}
\begin{tikzpicture}[scale=1.3]
\axes;
\person{\xB}{Bobbie};
\person{\xA}{{\color{red}Alex}};
\person{\xC}{Charlie};

\partnership{\xB}{\xA}{0}{\ymax}{0.2};
\partnership{\xA}{\xC}{0}{\ymax}{0.2};

\foreach \y in {1.6}   
   {\FailTrans{\xB}{\xA}{\y};};
\foreach \y in {0.3}
   {\FailTrans{\xA}{\xC}{\y};};

\InfectionStart{\xA}{0};
\SuccessTrans{\xC}{\xA}{0.3};
\InfectionStart{\xC}{0.3};
\SuccessTrans{\xA}{\xB}{1.6};
\InfectionStart{\xB}{1.6};
\end{tikzpicture}
\begin{tikzpicture}[scale=1.3]
\axes;
\person{\xB}{Bobbie};
\person{\xA}{Alex};
\person{\xC}{{\color{red}Charlie}};

\partnership{\xB}{\xA}{0}{\ymax}{0.2};
\partnership{\xA}{\xC}{0}{\ymax}{0.2};

\foreach \y in {1.6}   
   {\FailTrans{\xB}{\xA}{\y};};
\foreach \y in {0.3}
   {\FailTrans{\xA}{\xC}{\y};};

\InfectionStart{\xC}{0};
\SuccessTrans{\xC}{\xA}{0.3};
\InfectionStart{\xA}{0.3};
\SuccessTrans{\xA}{\xB}{1.6};
\InfectionStart{\xB}{1.6};
\end{tikzpicture}
\end{center}
\caption{\textbf{Scenarios for low transmission rates:} (top) Serial monogamy case in which each partnership transmits once.  (middle and bottom) Concurrent partnership case in which each partnership transmits once, with the order of transmission events differing in the two rows.  Without the shading to denote the partnerships, it would not be possible to infer which cases exhibit concurrency, so we expect similar population-scale outcomes.}
\label{fig:concurrency_display2}
\end{figure}

Investigating this closer in Fig.~\ref{fig:concurrency_display2}, we
see that indeed in the low transmission rate limit, we are unlikely to
be able to distinguish between concurrent and serial partnerships by
observing potential transmission events.  Once almost every
transmission event is to a different partner, the relevant detail is
simply the interval between transmission events.  The specific details
of how long a partnership lasts, or how many concurrent partnerships
exist are only marginally relevant.  As long as the probability of
transmitting twice to the same partner is negligible, the well-mixed population
assumption would provide the same prediction (see also the
``time-scale approximation" of~\cite{lloyd2004frequency} for more
discussion of this).  So the impact of concurrency when $\tau_1/\eta_1
\ll 1$ is negligible.

\subsubsection*{Equilibrium sizes}
We now turn to the equilibrium sizes which are predicted to be relatively insensitive to concurrency across much of parameter space.  Before investigating this, we highlight that many real populations are likely to have low partnership turnover rates ($\eta_1$), for which our model predicts the largest role of concurrency.

To explain the insensitivity of the equilibrium to concurrency, we note first that at equilibrium, an average infected individual causes one additional infection before leaving the population.  Consider now a serially monogamous population and a population with concurrency, both at equilibrium, and consider a single individual who becomes infected.

As noted in the introduction, by itself, concurrency does not cause an infected individual to be more or less likely to transmit to a newly acquired partner.  Rather, concurrency increases the probability an individual who is currently susceptible will later transmit to its currently susceptible partner.  

So for concurrency to affect the equilibrium level, when an individual becomes infectious he or she must have a non-negligible probability of infecting an existing partner \emph{at equilibrium}.  At equilibrium, we only expect one successful transmission (some other transmissions will fail because the recipient is already infected).  If the number of future partners is large compared to the number of current partners, we would expect that the one transmission is much more likely to go to a future partner than a current partner.

Put another way, if the risk to a susceptible individual from forming a new partnership with an already infected individual is much higher than the risk that an existing susceptible partner will become infected, then concurrency will not be a significant factor.

\subsubsection*{Transient growth}

We previously observed that concurrency can play a significant role in the early spread of disease even in scenarios where it has little effect on the long-term equilibrium.  In this section we first address why the effect of concurrency saturates.  Then we look at why the saturation occurs at higher levels of concurrency than for the equilibrium infection rates.  We note that for a given average number of transmissions, the early growth is higher if those transmissions are concentrated earlier in the infectious period~\cite{wallinga2007generation}.

Looking at the disease's perspective, early in an epidemic with a highly infectious disease, when an individual with many partners becomes infected, there will be rapid successful transmissions to many different partners.  In contrast, in the serial monogamy case, there will generally be a delay following a successful transmission because a partnership change is required before the next successful transmission: the partnership dynamics constrain the spread.  Thus in this scenario concurrency is expected to have an important impact on the early rate of spread.  

As concurrency increases, we are keeping the same number of transmissions per partnership, but only the first transmission in a partnership is successful.  By increasing concurrency, a larger fraction of the early transmissions an individual causes are the first transmission of the given partnership.  Increasing concurrency thus reduces the effect of local depletion of susceptibles.  At high enough concurrency the infected individual does not need to replenish its susceptible partner supply by changing partners.  At this point, further concurrency will have little impact.

To explain why the saturation of concurrency occurs at higher levels
for the growth rate than for the equilibrium level, we note that for
determining the equilibrium, what matters is how many infections an
individual causes, but the timing is not significant.  The early
growth however depends not just on how many infections are caused, but
also on how quickly those infections occur.  So the equilibrium is
most affected by the fact that concurrency can increase the number of
transmission chains, while the early growth is also affected by the
fact that concurrency increases the speed with which those
transmission chains are traced.  Additionally, in the early growth
regime, the increased number of transmission chains has a larger
effect than near equilibrium where a non-negligible fraction of those
additional chains are blocked by existing infection.

\subsubsection*{Implications for intervention design}
A major source of controversy about designing interventions to reduce
concurrency is that it necessarily takes resources away from other
interventions, and so those who question the magnitude of its effect
understandably question the wisdom of implementing interventions.

Concurrency reduction would reduce transmissions early in the
infectious period.  In contrast, many interventions under
consideration require identifying infected individuals and reducing
their probability of onwards transmission.  As these interventions are
scaled up, a larger proportion of transmissions will be from more
recently infected individuals increasing the relative value of
interventions preventing transmissions from recently infected
individuals.  Thus concurrency reducing interventions may be a good
complement to other interventions or a good followup once these other
interventions are widely implemented.

We additionally note that our model raises the possibility that concurrency may have played a role in the rapid growth of the HIV epidemic in some regions, but that the role of concurrency in determining the resulting level of infection may have saturated.  Thus, now that the epidemic is well-established, moderate reductions in concurrency might not lead to a rapid decay in the epidemic.  This raises the threshold required for concurrency reduction to be effective: for a wide range of parameters reducing concurrency from a high level to a moderate level has much less impact on the epidemic than reducing concurrency from a moderate level to a low level.  Thus concurrency-reducing interventions in well-established epidemics may be most effective in lower-concurrency settings.

An important setting which is not investigated in our model is the case in which a significant fraction of the population does not engage in concurrency.  For these individuals, we would expect concurrent relationships of their partners to be a major source of their infection risk.  Thus an intervention which encourages non-concurrent individuals to ensure that they partners who are also non-concurrent may well be successful.  Our model could be used to test this with minor modifications.

\subsubsection*{Caveats}

There are a number of caveats of our study that must be highlighted to avoid overinterpreting these results.  
Understanding these limitations and why they might arise gives
guidance on when we should expect concurrency to be important.

\begin{itemize}
\item \textbf{Acute Phase:} Before mounting an immune response, an
  individual's viral load is several orders of magnitude larger than
  after the immune response develops.  During this early phase
  infectiousness is dramatically
  increased~\cite{pilcher:acute_hiv,brenner:acute,pilcher2007amplified,chibo:acute,pinkerton2008probability}.
  If the individual has multiple partnerships, then many more infections can happen in this phase than would be seen if the individual were only in contact with its infector~\cite{eaton2011concurrent}.

\item \textbf{Impact of future interventions} As ``treatment as
  prevention" or other interventions are implemented, it is likely
  that later partners will be at lower risk because treatment will
  reduce an individual's infectiousness.  This will increase the role
  of transmissions occurring early in an infectious period, increasing
  the relative role of concurrency.

\item \textbf{Heterogeneous degree:} Some people have many more partnerships than others~\cite{liljeros}.  They generally become infected sooner, and in turn transmit to more individuals.  Even if many individuals do not engage in concurrent relationships, if there a few with many concurrent partners, the effects may still be present.  This provides an opportunity to reduce disease transmission through an intervention that encourages those without concurrent relationships to ensure their partners also do not have concurrent relationships.

\item \textbf{Temporal behavior changes:} If the disease dynamics are driven by individuals having periodic high-risk episodes between long-term relationships, then the assumptions of this model are invalid.  To correct this, the model must be adapted to allow for periods of high risk behavior, for example when a partnership ends.

\item \textbf{Age structure:} If there is age-structure in the contact
  patterns, different effects may be seen.  For example, we might
  think of the younger cohort as a population which has not yet been
  invaded by infection.  In this case, the results about early growth
  as the disease invades this subpopulation may be more relevant than
  our model predicts.  Reducing concurrency could be expected to play an important role in slowing the invasion of this younger cohort.

\item \textbf{Coital dilution:} We have assumed that the transmission rate scales such that individuals have effectively the same total number of sexual acts regardless of their number of partners.  This dilutes the number of acts per partnership, and to address this we extended the partnerships.  This allows us to isolate the effect of concurrency from the effect of frequency of sexual acts.  However, if concurrent relationships are associated with more (or less) frequent sexual acts, then the conclusions we reach here may not be valid.  To correct for this, we would need to appropriately weight the transmission rates based on the number of concurrent relationships each partner has~\cite{althaus2015sex}.
\end{itemize}

\section*{Conclusions}

We have derived an analytic model which accurately reproduces simulated SI epidemics in a population with concurrent relationships and demographic turnover.  We use this model to isolate the role of concurrency in the spread of a disease such as HIV.  

Although the model is highly simplistic, it can be generalized to
incorporate more realism, and it can be used to help us understand
important features of the role of concurrency in HIV spread.  We see first that the impact of concurrency on the equilibrium size of SI epidemics can saturate.  Consequently interventions targeting concurrency may have little impact unless they come close to eliminating concurrent relationships.  
However, we see a more significant role for concurrency in determining
the early growth rate.  As concurrency increases, the early growth is
increased, and the effect saturates at higher concurrency levels than for the epidemic size.  Thus reducing concurrency is likely to have more impact on early growth than on the final equilibrium.

An important additional observation from our analysis is that we can
make significant progress by focusing on the ``disease-eye'' view of
the population~\cite{adams2013sex}.  Using this, we have been able to explain why the
numerical predictions of the model follow from the assumptions, and
also gain insight into conditions under which we could expect our
predictions to be robust to real-world conditions.

Our model is intended as a framework for developing more realistic models.  Our goal has been to provide this framework and clearly demonstrate that it is possible to use analytic models to explore disease spread in populations with concurrent relationships with demographic turnover.  The predictions our model has provided are true for the simplistic assumptions made.  More careful models will be needed to identify conditions under which interventions targeting concurrency will be effective.  These models will need to incorporate additional effects such as the acute phase of infection and more realistic information about degree distributions and correlations.  

Our modeling introduces new issues which have not previously been considered in the concurrency discussion.  In particular, even if a population has significant concurrency and even if that concurrency played a major role in the establishment and growth of the HIV epidemic in some population, it is not guaranteed that concurrency plays an important role in the current levels of infection.  Thus although concurrency may cause an epidemic to grow quickly to its equilibrium, it is not clear that once the population reaches equilibrium reducing concurrency would significantly affect the equilibrium.

Regardless of the magnitude of the current impact of concurrency, it is likely that interventions such as TasP will disproportionately reduce transmissions caused later in the infectious period compared to transmissions caused earlier in the infectious period. As this happens, the relative impact of concurrency will increase.

\section*{Supporting Information}

Our primary goal in the Supporting Information is to derive the governing equations.  Although in the main text we assume all individuals have the same number of concurrent partners, in our derivation here we allow different individuals within the same population to have a different number of concurrent partners as long as all partnerships have the same transmission probability and typical duration.  These assumptions could be modified and a number of other complexities added to the model we develop here, but we do not attempt this now.

\subsection*{Stochastic population and disease model}
We now describe the stochastic rules we assume govern the population and disease dynamics.  We use a discrete-time model.  We begin with the population dynamics in the absence of disease.  At each time step, $N\mu$ individuals enter the population, and each individual has probability $\mu$ to independently leave the population.  This leads to an equilibrium population size of $N$, but with variation around this value.

Each individual $u$ has a constant number of partners $k_u$ which is
assigned independently to $u$ when $u$ enters the population.  $P(k)$
gives the probability that $k_u=k$.  We think of $u$ as having $k_u$
``stubs'' (also called ``binding sites''
by~\cite{leung:demographic,leung:disease}).  The stubs pair with stubs
from other individuals to form partnerships.  When a partnership ends
the two newly freed stubs join with other free stubs to form new
partnerships.  We assume that individuals immediately replace their
partners so that at the start of each time step all individuals have a
full set of partners (\cite{miller:ebcm_overview} has discussion of how to include more complicated partnership  dynamics.).

We are interested in the epidemic timescale, which is longer than the individual's active period.  So we must include ``birth'' and ``death'' or equivalently immigration and emigration.

We begin with a fraction $\rho$ of the population randomly infected. In each time step, multiple events can happen.  Since the order of events can matter (a partnership cannot transmit after it ends), we provide a consistent order, shown in figure~\ref{fig:order_of_events}.  First, infected individuals transmit to their susceptible partners independently with probability $\tau$.  Second, individuals may ``die'' (or leave the population) independently with probability $\mu$.  Third, $\mu N$ new individuals are added to the population (so the average number present is $N$) and assigned stubs.  Fourth, each remaining partnership breaks with probability $\eta$.  Finally, the unpaired stubs form new partnerships, subject to the constraint that old partnerships are not reformed and individuals do not join to themselves.  In simulations, these constraints are occasionally not satisfied, in which case the corresponding individuals wait a time step before attempting to form new partnerships.  In a large population, the impact of this failure is negligible, and for our analytic equations below, we can assume that they are satisfied.

\subsection*{Equation Derivation}
We now derive the discrete-time equations presented in the main text as well as a continuous-time version.  These equations govern the large-population limit of our model.
\subsubsection*{Preliminaries}
It will be useful to define the function 
\[
\psi(x) = \sum_k P(k) x^k
\]
to be the probability generating function of the degree distribution.  It has some important properties: $\psi(1)=\sum P(k)1^k = 1$, \ $\psi'(1) = \sum k P(k)1^{k-1} = \ave{K}$ where $\ave{\cdot}$ denotes the mean of the random variable.

Our derivation is based on~\cite{miller:ebcm_overview}.  We review the concept of a ``test individual'' (effectively equivalent to the \emph{cavity state} of~\cite{karrer:message}).  We start with the assumption that the population-scale dynamics are deterministic in the large population limit.  A direct consequence of this assumption is the observation that the probability a randomly selected individual has a given status equals the proportion of the population with that status.

Although in the asymptotic limit, they have the same value, calculating the probability a random individual has a given status turns out to be simpler than calculating the proportion of individuals in each state.  This is because of a simplification that results from the observation that the probability a single randomly chosen individual $u$ has a given status is not affected if we prevent $u$ from infecting any other individuals (Although it is not necessary here, it may be helpful to recognize that that the assumption the stochastic process exhibits deterministic population-scale dynamics means that a change of out come for a vanishingly small fraction of events does not alter the population-scale dynamics.).  If we prevent $u$ from transmitting to its partners, then the status of its partners become independent of one another.

Guided by this, we define a \emph{test individual} to be an individual $u$ chosen uniformly at random from the population and prevented from transmitting infection.  We have the following sequence of questions which have identical answers if the dynamics are deterministic:  Given the initial proportion infected $\rho$,
\begin{enumerate}
\item What fraction of individuals are susceptible or infected at time $t$?
\item What is the probability a random individual is susceptible or infected at time $t$?
\item What is the probability a randomly chosen test individual is susceptible or infected at time $t$?
\end{enumerate}
The first two equations have the same answer because we assume $\rho N$ is large enough that the dynamics may be treated as deterministic. The last two equations have the same answer because preventing a single individual $u$ from transmitting does not affect \emph{its} probability of changing status (we highlight that we are \emph{not} asking what proportion of nodes are in each state once $u$ is prevented from transmitting).

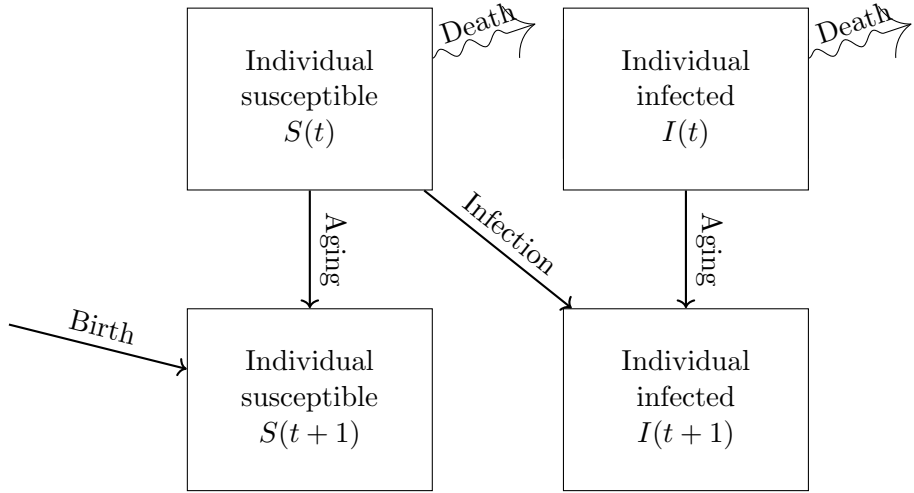
\begin{figure}
\begin{center}
\begin{tikzpicture}
\node [box] (It) at (5,0) {\parbox{3cm}{\begin{center}Individual infected\\ $I(t)$\end{center}}};
\node [box] (St) at (0,0) {\parbox{3cm}{\begin{center} Individual susceptible\\ $S(t)$\end{center}}};
\node [box] (It) at (5,0) {\parbox{3cm}{\begin{center}Individual infected\\ $I(t)$\end{center}}};
\node [box] (Stp1) at (0,-4) {\parbox{3cm}{\begin{center}Individual susceptible\\ $S(t+1)$\end{center}}};
\node [box] (Itp1) at (5,-4) {\parbox{3cm}{\begin{center}Individual infected\\ $I(t+1)$\end{center}}};

\path [thickpath] (St) -- (Itp1) node [midway, sloped, above] {Infection};
\path [thickpath] (St) -- (Stp1) node [midway, sloped, above] {Aging};
\path [thickpath] (It) -- (Itp1) node [midway, sloped, above]{Aging};
\path [decay] (St) -- ($(St)+(3,1)$) node [midway, sloped, above] {Death};
\path [decay] (It) -- ($(It)+(3,1)$) node [midway, sloped, above] {Death};
\path [thickpath] ($(Stp1)+(-4,1)$) -- (Stp1) node [midway, sloped, above] {Birth};
\end{tikzpicture}
\end{center}
\caption{Basic transitions of individuals from time $t$ to $t+1$.  At each time step, death occurs with probability $\mu$.  At population equilibrium, the proportion of the population that has age $0$ is thus the number born divided by the total population, that is, $\mu= b/N$.  The class $S$ can be subdivided based on how long an individual has been in the population.  Note that deaths are exactly balanced by births (at population equilibrium).}
\label{fig:basicSI}
\end{figure}

\begin{figure}
\begin{center}
\begin{tikzpicture}[scale=0.8]
\pgfmathsetmacro{\x}{3.3}
\node [box] (S0) at (0,0) {$\mu s(t,0)$};
\node [] (Stlabel) at (\x, 1.) {$S(t)$};
\node [] (dots) at (0.75*\x,0) {$\cdots$};
\node [box] (St) at (1.5*\x,0) {$\mu(1-\mu)^{a_u}s(t,a_u)$};
\node [] (2dots) at (2.25*\x,0) {$\cdots$};\node [box] (S0) at (0,0) {$\mu s(t,0)$};
\node [box] (It) at (4.*\x, 0) {$I(t)$};
\node [box] (S0tp1) at (0,-2) {$\mu s(t+1,0)$};
\node [box] (Stp1) at (2*\x,-2) {$\mu(1-\mu)^{a_u+1}s(t+1,a_u+1)$};
\node [] (Stp1label) at (1*\x,-3) {$S(t+1)$}; 
\node [] (3dots) at (0.75*\x, -2) {$\qquad\cdots$};
\node [] (4dots) at (3*\x, -2) {$\cdots$};
\node [box] (Itp1) at (4.*\x, -2) {$I(t+1)$};
\coordinate (join) at (3*\x,-1);
\begin{pgfonlayer}{background}
\node [draw, fit=(S0)(2dots)(Stlabel), fill = gray!10] (S1) {};
\node [draw, fit = (S0tp1)(Stp1)(Stp1label)(3dots)(4dots), fill =
gray!10] (S2) {};
\end{pgfonlayer}
\path [thickpath] (St) -- (Stp1) {};
\path [thickpath] (S0) -- (3dots) {};
\path [->, out=330, in = 180] (St) edge node {} (join);
\path [->, out = 340, in = 180] (S0) edge node {} (join);
\path [->, out = 0, in = 150] (join) edge node {} (Itp1);
\path [thickpath] (It) -- (Itp1) {};
\path [decay] (St) -- ($(St) + (2,2)$); 
\path [decay] (S0) -- ($(S0) + (2,2)$); 
\path [decay] (It) -- ($(It) + (2,2)$); 
\path [thickpath] ($(S0tp1)+(-2,1)$) -- (S0tp1) {};
\end{tikzpicture}

\end{center}
\caption{The subdivisions of class $S$.  The variable $s(t,a_u)$
  denotes the probability that an age $a_u$ individual is susceptible.
  Thus $s(t,0)=1$.  The number of individuals in each age class is
  $\mu(1-\mu)^{a_u}$, representing the fact that a proportion $\mu$
  are born in a given time step, and of these $(1-\mu)^{a_u}$ survive to step $a_u$.  The arrows show the fluxes out of the compartment corresponding to susceptible individuals of age $a_u$.  There is no need to subdivide the $I(t)$ compartment.  Rather than calculating the flux along each arrow, we will derive explicit expressions for $s(t,a_u)$.}
  \label{fig:subdividedS}
\end{figure}
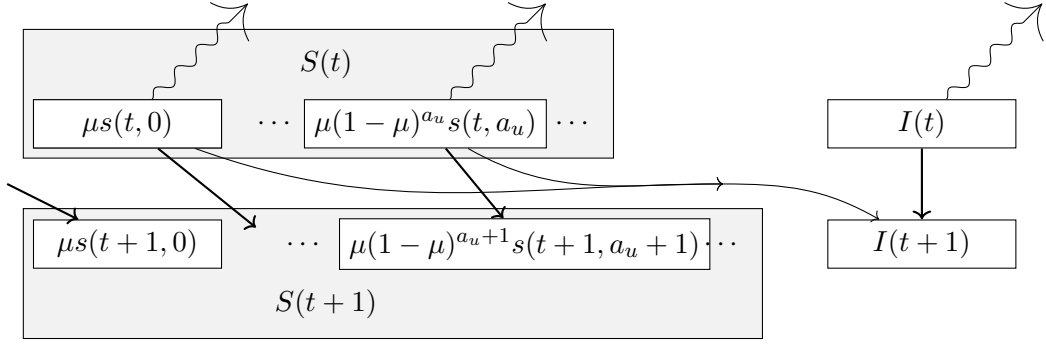

We start our calculations with the goal of finding $S(t)$ and $I(t)$ using figure~\ref{fig:basicSI}.  The processes such as death and aging and birth are relatively straightforward to model.  However, the probability an individual becomes infected in a time step is not independent of the individual's age: For example, an individual who has been in the population longer will have a different distribution of partnerships from someone who has only recently joined.  Consequently we must subdivide $S(t)$ based on individual age $a_u$.  
These subdivisions are shown in figure~\ref{fig:subdividedS}.  At equilibrium, the proportion of the population which is age $0$ is $\mu$ (which equals $b/N$), and the probability that such an individual is susceptible is $s(t,0)=1$.  At each subsequent time step, these aging individuals are removed with probability $\mu$, and so the proportion of the population with age $a_u$ is $\mu(1-\mu)^{a_u}$.  The probability that such an individual is susceptible is defined to be $s(t,a_u)$. To find $s(t,a_u)$, we turn to $\Theta(t,a_u)$, the probability that a stub belonging to an age $a_u$ test individual $u$ has never been involved in a transmission to $u$.  Once we know that, then the probability a test individual of age $a_u$ and $k_u$ partners is susceptible at time $t$ is $\Theta(t,a_u)^{k_u}$.  Averaging this over the entire population of age $a_u$ individuals the probability an age $a_u$ individual is susceptible is 
\[
s(t,a_u) = \begin{cases}
\psi(\Theta(t,a_u)) & a_u < t\\
(1-\rho)\psi(\Theta(t,a_u)) & a_u \geq t
\end{cases}
\]
where we recall $\sum_k P(k) x^k = \psi(x)$, and the $1-\rho$ factor in the second term accounts for the fact that the individual would be infected at $t=0$ with probability $\rho$.

The fraction susceptible is thus
\[
S(t) = \mu \sum_{a_u=0}^\infty (1-\mu)^{a_u} s(t,a_u)
\]
The probability of being infected is
\[
I(t) = 1-S(t)
\]
We must now derive an expression for $s(t,a_u)$.

\begin{figure}
\begin{center}
\begin{tikzpicture}
\node [box] (PhiIt) at (5,0)
{\parbox{3cm}{\begin{center}Partner infected, stub has not transmitted\\ $\Phi_I(t,a_u)$\end{center}}};
\node [box] (PhiSt) at (0,0) {\parbox{3cm}{\begin{center}
      Partner susceptible, stub has not transmitted\\ $\Phi_S(t,a_u)$\end{center}}};
\node [box] (PhiIt) at (5,0) {\parbox{3cm}{\begin{center}Partner infected, stub has not transmitted\\ $\Phi_I(t,a_u)$\end{center}}};
\node [box] (PhiStp1) at (0,-4) {\parbox{3cm}{\begin{center} $\Phi_S(t+1,a_u+1)$\end{center}}};
\node [box] (PhiItp1) at (5,-4) {\parbox{3cm}{\begin{center} $\Phi_I(t+1,a_u+1)$\end{center}}};
\node [box, minimum width = 1cm] (1mt) at (9,0.5) {\parbox{1.75cm}{\begin{center}$1-\Theta(t,a_u)$\end{center}}};
\node [box, minimum width = 1cm] (1mttp1) at (9,-4) {\parbox{1.75cm}{\begin{center}$1-\Theta(t+1,a_u+1)$\end{center}}};
\node [] (thetatlabel) at (3.1,1.5) {$\Theta(t,a_u)$};
\node [] (thetatp1label) at (2.5,-5) {$\Theta(t+1,a_u+1)$};
\begin{pgfonlayer}{background}
\node [draw, fit=(PhiSt)(PhiIt)(thetatlabel), fill = gray!10] (ThetaT) {};
\node [draw, fit=(PhiStp1)(PhiItp1)(thetatp1label), fill = gray!10] (ThetaTp1) {};
\end{pgfonlayer}
\coordinate (join) at (1.7,2.7);
\coordinate (break) at (-2.5,-0.8);

\path [thickpath] (PhiSt) -- (PhiItp1) node [midway, sloped, above]
{Infection of partner};
\path [thickpath] (PhiSt) -- (PhiStp1) node [midway, sloped, above] {Aging};
\path [thickpath] (PhiIt) -- (PhiItp1) node [midway, sloped, above]{Aging};
\path [thickpath] (PhiIt) -- (1mttp1) node [midway, sloped, above] {Transmission};
\path [thickpath] (1mt) -- (1mttp1) node [midway, sloped, above]
{Aging};
\path [->, above, out=40, in = 0] (PhiIt) edge node {} (join);
\path [->, above, out=20, in = 0] (PhiSt) edge node {} (join);
\path [->, above, sloped, out=180,in=90, pos=0.2](join) edge node {new partnership
forming} (break);
\path [->, above, out=270, in=160] (break) edge node {} (PhiStp1);
\path [->, above, out=270, in=160] (break) edge node {} (PhiItp1);
\end{tikzpicture}
\end{center}
\caption{Basic transitions of partnerships from time $t$ to $t+1$ (from the perspective of an age $a_u$ test individual who does not die at time $t$).  At each time step the partnership ends with probability $\mu + \eta$. .  The class $\Phi_S$ can be subdivided based on how long a partnership has existed.}
  \label{fig:theta}
\end{figure}
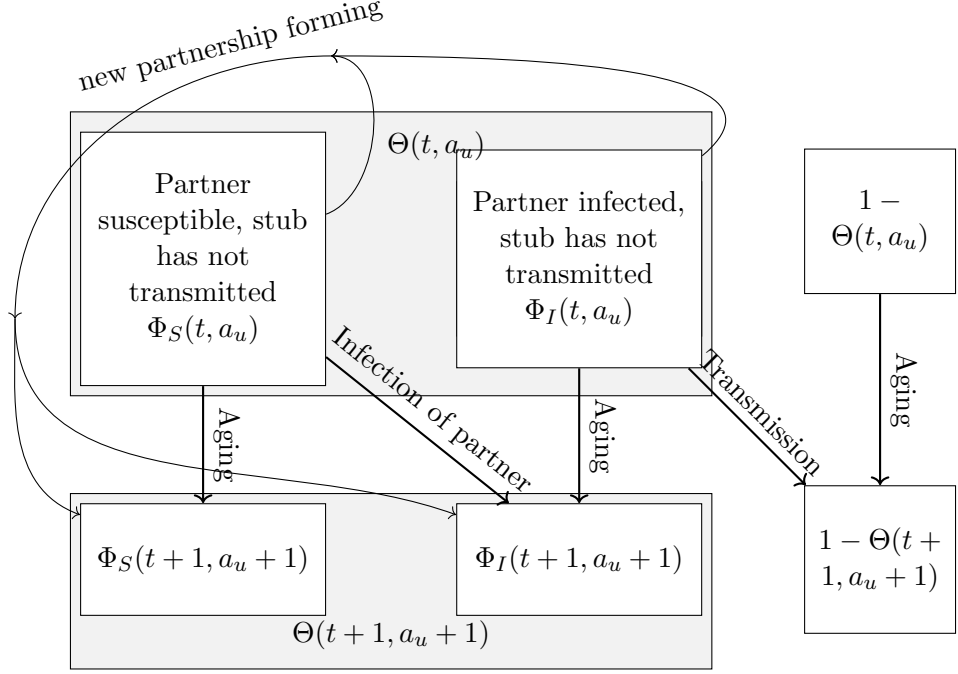

\begin{figure}
\begin{center}
\begin{tikzpicture}[scale=0.9]
\pgfmathsetmacro{\x}{2.8}
\node [box] (S0) at (0,0) {$p_b \phi_S(t,a_u,0)$};
\node [] (PhiStlabel) at (\x, 1.) {$\Phi_S(t,a_u)$};
\node [] (dots) at (0.65*\x,0) {$\cdots$};
\node [box] (phiSt) at (1.65*\x,0) {$p_b(1-p_b)^{a_u}\phi_S(t,a_u,a_e)$};
\node [] (2dots) at (2.75*\x,0) {$\cdots$};
\node [box] (phiS0) at (0,0) {$p_b \phi_S(t,a_u,0)$};
\node [box] (phiStp1e0) at (0, -2) {$p_b \phi_S(t+1,a_u+1,0)$};
\node [box] (phiStp1) at (2*\x,-2) {$p_b(1-p_b)^{a_u}\phi_S(t+1,a_u+1,a_e+1)$};
\node [] (PhiStp1label) at (1.5*\x,-3) {$\Phi_S(t,a_u+1)$}; 
\node [] (3dots) at (0.8*\x, -2) {$\cdots$};
\node [] (4dots) at (3.25*\x, -2) {$\cdots$};
\node [box] (PhiIt) at (4.1*\x, 0) {$\Phi_I(t,a_u)$};
\node [box] (PhiItp1) at (4.1*\x, -2.5) {$\Phi_I(t+1,a_u+1)$};
\begin{pgfonlayer}{background}
\node [draw, fit=(S0)(2dots)(PhiStlabel), fill = gray!10] (S1) {};
\node [draw, fit = (phiStp1)(PhiStp1label)(phiStp1e0)(4dots), fill = gray!10] (S2) {};
\end{pgfonlayer}
\coordinate (join) at (1.25,2);
\coordinate (break) at (-2.,0.5);

\path [thickpath] (phiSt) -- (phiStp1) {};
\path [thickpath] (phiSt) -- (PhiItp1) {};
\path [->, above, out=20, in = 0] (phiSt) edge node {} (join); 
\path [->, above, out=20, in = 0] (phiS0) edge node {} (join); 
\path [->, above, sloped, out=180,in=90](join) edge node {} (break);
\path [->, above, out=270, in=140] (break) edge node {} (phiStp1e0);
\path [-, above, out=270, in=180] (break) edge node {} (2.5,-1) ;
\path [->, above, out = 0, in=145] (2.5,-1) edge node {} (PhiItp1);
\path [thickpath] (PhiIt) -- (PhiItp1) {};
\end{tikzpicture}
\end{center}
\caption{The subdivisions of the $\Phi_S$ compartment.  Arrows show
  the fluxes out of the subcompartment corresponding to partnerships of age $a_e$.  Rather than calculating the fluxes along each arrow, we will explicitly calculate $\phi_S(t,a_u,a_e)$.}
  \label{fig:phiS}
\end{figure}
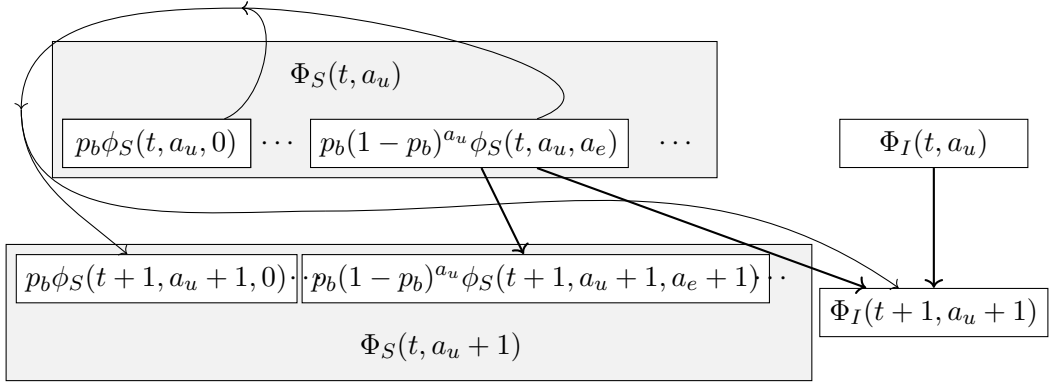

The focus of our calculations is on determining $\Theta(t, a_u)$.  As
a boundary condition we have 
\[
\Theta(t,0) = 1
\]
stating that when an individual is first introduced, it has not yet received
any infection.  Similarly we have the initial condition
\[
\Theta(0,a_u)=1
\]
as well, stating that prior to the disease introduction, no
transmissions have occurred.  Looking at figure~\ref{fig:theta}, we see that the change in $\Theta$ is from transmissions which occur with probability $\tau \Phi_I$.  So the change in $\Theta$ in a time step is
$-\tau \Phi_I$ where $\Phi_I$ is the probability that the stub has not
previously brought infection to $u$ and connects to an infected
partner at the start of the time step.  So we have
\[
\Theta(t,a_u) = \Theta(t-1,a_u-1) - \tau \Phi_I(t-1,a_u-1)
\]
However to do this calculation we require $\Phi_I(t,a_u)$ which is still unknown.  We can
shift our unknown from $\Phi_I$ to $\Phi_S$ (the probability the stub
has not transmitted to $u$ and currently connects to a susceptible
partner) by using
\[
\Phi_I = \Theta-\Phi_S \, .
\]

As in calculating $S$, to calculate $\Phi_S$, we turn it into a sum, following figure~\ref{fig:phiS}.
The probability that a partnership created when $u$ joined still
exists is $(1-p_b)^{a_u}$.  The probability that a partnership has
some smaller age $a_e$ is $p_b(1-p_b)^{a_e}$.  So
\[
\Phi_S =  (1-p_b)^{a_u}\phi_S(t,a_u,a_u) + p_b\sum_{a_e=0}^{a_u-1}(1-p_b)^{a_e} \phi_S(t,a_u,a_e)
\]
where $\phi_S(t,a_u,a_e)$ is the probability that a stub belonging to
an age $a_u$ individual that is part of an age $a_e$ partnership has not
transmitted by time $t$ and $p_b$ is the probability that a stub is
freed to find a new partnership  (either by death of the partner, or termination of
the partnership).  The one term outside the sum represents the fact
that when the individual first enters the population the stub
definitely forms a partnership.

We now find $\phi_S(t,a_u,a_e)$.  If the partnership formed when $u$ was born ($a_e=a_u$) then this is simply the probability the partner $v$ is susceptible given that $v$ has an age $a_u$ partnership with $u$, which we denote $\chi(t,a_u)$.  However, if the partnership formed after $u$ was born ($a_e<a_u$), then $\phi_S(t,a_u,a_e)$ is the probability 
$\Theta(t-a_e,a_u-a_e)$ that the stub
was not responsible for transmitting infection to individual $u$ prior
to the current partnership forming times $\chi(t,a_e)$.  As
$\Theta(t-a_u,0)=1$ these coincide when $a_u=a_e$, so we can write
\[
\phi_S(t,a_u,a_e) = \Theta(t-a_e,a_u-a_e) \chi(t,a_e)
\]

We now find $\chi(t,a_e)$ similarly to $s(t,a_u)$.  It is 
\[
\chi(t,a_e)=\sum_{A_v=a_e}^\infty P(a_v=A_v|a_e) P(v \text{ susceptible}|a_v=A_v) \, .
\]
If $a_e\geq t$, then we know that $v$ was born either when the disease was introduced or earlier.  Thus no previous partnership could have transmitted to $v$.  if we assume $a_v=A_v \geq a_e$, then the probability $v$ is susceptible is the probability that it escaped infection when the disease was introduced $1-\rho$ times the probability that it has not been infected by any other partners.  Because of how $v$ is selected (it is $u$'s partner), $v$ is likely to have a higher degree than a randomly selected individual.  The probability $v$ has degree $k_v=k$ is $k P(k)/\ave{K}$.  So the probability $v$ is susceptible given $A_v$ is  $(1-\rho)\sum_k [k P(k)/\ave{K}] \Theta(t,A_v)^{k-1} = (1-\rho)\psi'(\Theta(t,A_v))/\ave{K}$.  Thus for $a_e \geq t$ we have
\begin{align*}
\chi(t,a_e) &= \sum_{A_v=a_e}^\infty  P(a_v=A_v|a_e) P(v \text{ susceptible} | a_v=A_v)\\
&= \sum_{a_v=a_e}^\infty \mu(1-\mu)^{a_v-a_e}(1-\rho) \frac{\psi'(\Theta(t,a_v))}{\ave{K}} \qquad a_e \geq t
\end{align*}
For $a_e<t$ there are three important cases to consider based on whether the partner was born before or at the same time that the partnership was formed and whether the partner was born before or after the disease was introduced.  
\begin{itemize}
\item If the partnership formed when $v$ was born then $A_v=a_e$ for which $P(a_v=a_e|a_e) = (1-P_e)$ and $P(v \text{ susceptible} | a_v=a_e)=\sum_k [k P(k)/\ave{K}] \Theta(t,a_e)^{k-1} = \psi'(\Theta(t,a_e))/\ave{K}$, which measures the probability that another partner of $v$ has not transmitted to $v$.  
\item If $v$ was born before the partnership formed but after the disease was introduced then $a_e<A_v<t$ and $P(a_v=A_v|a_e) = P_e \mu(1-\mu)^{A_v-a_e-1}$. Although $u$ has not transmitted to $v$, it is possible that a previous partner of $v$ that was eventually replaced by $u$ did.  Thus the probability $v$ is susceptible is $\Theta(t-a_e,A_v-a_e)\psi'(\Theta(t,A_v))$.
\item If $v$ was born before the disease was introduced, then $A_v \geq t$.  We again have $P(a_v=A_v|a_e) = P_e \mu(1-\mu)^{A_v-a_e-1}$, but there is an extra factor of $1-\rho$ in the probability $v$ is susceptible.  $P(v \text{ susceptible} | A_v) = (1-\rho)\Theta(t-a_e,A_v-a_e)\psi'(\Theta(t,A_v))$.
\end{itemize}
So for $a_e<t$ we have
\begin{align*}
\chi(t,a_e) &= \sum_{A_v=a_e}^\infty P(a_v=A_v|a_e) P(v \text{ susceptible}|a_v)\\
&= P(a_v=a_e|a_e)P(v \text{ susceptible} | a_v=a_e)\\
&\quad+ \sum_{_v=a_e+1}^{t-1} P(a_v=A_v|a_e) P(v \text{ susceptible}|a_v=A_v) \\
&\quad+\sum_{A_v=t}^{\infty} P(a_v=A_v|a_e) P(v \text{ susceptible}|a_v=A_v)\\
&= (1-P_e) \frac{\psi'(\Theta(t,a_e))}{\ave{K}} \\
&\quad+ P_e\mu\sum_{a_v=a_e+1}^{t-1}(1-\mu)^{a_v-a_e-1} \Theta(t-a_e,a_v-a_e)\frac{\psi'(\Theta(t,a_v))}{\ave{K}}\\
&\quad + P_e\mu (1-\rho)\sum_{a_v=t}^{\infty}(1-\mu)^{a_v-a_e-1} \Theta(t-a_e,a_v-a_e)\frac{\psi'(\Theta(t,a_v))}{\ave{K}}\\
\end{align*}

\subsubsection*{Simplification for $a_u>t$}

We claim that the value of $\Theta(t,a_u)$ is the same for all $a_u\geq t$.  This follows from the fact that at $t=0$ all the values are $1$.  By inspecting the equations for the evolution of $\Theta$, we see that if we assume $\Theta(t,a_u)$ is the same for all $a_u \geq t$, then the change in $\Theta$ is also the same.  Thus we can assume $\Theta(t,a_u)=\Theta(t,t)$ if $a_u>t$.  This argument would break down if partnership formation were affected by age differences.  

Among the resulting simplifications is the observation that for $a_e \geq t$, the expression for $\chi(t,a_e)$ simplifies to $(1-\rho) \psi'(\Theta(t,t))/\ave{K}$.

\subsection*{Governing Equations}
Our full system of equations becomes
\begin{align*}
S(t) &= \mu \sum_{a_u=0}^{\infty} (1-\mu)^{a_u} s(t,a_u)\\
s(t,a_u) &= \begin{cases} \psi(\Theta(t,a_u)) & a_u < t\\
     (1-\rho)\psi(\Theta(t,t)) & a_u \geq t
     \end{cases}\\
I(t) &= 1-S(t)\\
\Theta(t,0)&=1\\
\Theta(0,a_u)&=1\\
\Theta(t,a_u) &= \Theta(t-1,a_u-1) - \tau \Phi_I(t-1,a_u-1) \qquad t, a_u\geq 1\\
\Phi_I(t,a_u) &= \Theta(t,a_u)-\Phi_S(t,a_u) \\
\Phi_S(t,a_u) &= (1-p_b)^{a_u} \phi_S(t,a_u,a_u) + p_b \sum_{a_e=0}^{a_u-1} (1-p_b)^{a_e}\phi_S(t,a_u,a_e)\\
\phi_S(t,a_u,a_e) &= \Theta(t-a_e, a_u-a_e) \chi(t,a_e)\\
\chi(t,a_e) &= \begin{cases}
  (1-\rho)\frac{\psi'(\Theta(t,t))}{\ave{K}} & a_e \geq t\\[12pt]
(1-P_e) \frac{\psi'(\Theta(t,a_e))}{\ave{K}}\\
+ P_e \mu \sum_{a_v=a_e+1}^{t-1} (1-\mu)^{a_v-a_e-1}\Theta(t-a_e,A_v-a_e)\frac{\psi'(\Theta(t,a_v))}{\ave{K}} & a_e < t\\
+ P_e (1-\rho)\Theta(t-a_e,t-a_e)\frac{\psi'(\Theta(t,t))}{\ave{K}} (1-\mu)^{t-a_e-1}
\end{cases}
\end{align*}
We can derive a differential equations version of this by 
treating the time step as $\Delta t$ rather than $1$ and assuming that the event probabilities are all proportional to $\Delta t$.  Then taking $\Delta t \to 0$ yields differential equations.  We will use $\hat{\mu} = \lim_{\Delta t \to 0} \mu/\Delta t$ and similarly define other variables.

In the continuous time case, we find
\begin{align*}
S(t) &= \mu \int_0^\infty e^{-a_u \mu} s(t,a_u) \, \mathrm{d}a_u\\
s(t,a_u) &= \begin{cases}
\psi(\Theta(t, a_u)) & a_u < t\\
(1-\rho) \psi(\Theta(t,t)) & a_u \geq t
\end{cases}\\
I(t) &= 1-S(t)\\
\Theta(t,0) &= 1\\
\Theta(0, a_u) &= 1\\
\left(\pd{}{t} + \pd{}{a}\right) \Theta(t, a_u) &= - \tau \Phi_I(t,a_u)\\
\Phi_I(t, a_u) &= \Theta(t, a_u) - \Phi_S(t, a_u)\\
\Phi_S(t, a_u) &= e^{-p_b a_u} \phi_S(t,a_u,a_u) + p_b \int_0^{a_u} e^{-p_b a_v}\phi_S(t,a_u,a_e) \, \mathrm{d}a_e\\
\phi_S(t,a_u,a_e) &= \Theta(t-a_e, a_u-a_e)\chi(t,a_e)\\
\chi(t,a_e) &= \begin{cases}
(1-\rho) \frac{\psi'(\Theta(t,t))}{\ave{K}} & a_e \geq t\\[12pt]
(1-P_e) \frac{\psi'(\Theta(t,a_e))}{\ave{K}}\\ 
+ P_e \mu \int_{a_e}^t e^{-\mu(A_v-a_e)} \Theta(t-a_e, A_v-a_e) \frac{\psi'(\Theta(t,A_v))}{\ave{K}} \, \mathrm{d}A_v & a_e<t\\
+ P_e(1-\rho)\Theta(t-a_e,t-a_e) \frac{\psi'(\Theta(t,t))}{\ave{K}} e^{-\mu(t-a_e)}
\end{cases}
\end{align*}
The simplest numerical method to solve this system of equations would discretize by age and apply an Euler method, which corresponds to solving the discrete-time equations above.

\bibliographystyle{plain}
\bibliography{birthdeath}

\providecommand{\noopsort}[1]{}
\begin{thebibliography}{10}

\bibitem{adams2013sex}
Jimi Adams, James Moody, and Martina Morris.
\newblock Sex, drugs, and race: how behaviors differentially contribute to the
  sexually transmitted infection risk network structure.
\newblock {\em American journal of public health}, 103(2):322--329, 2013.

\bibitem{althaus2015sex}
Christian~L Althaus, Marc Choisy, and Samuel Alizon.
\newblock How sex acts scale with the number of sex partners: evidence from
  chlamydia trachomatis data and implications for control.
\newblock {\em PeerJ PrePrints}, 3:e1821, 2015.

\bibitem{brenner:acute}
Bluma~G Brenner, Michel Roger, Jean-Pierre Routy, Daniela Moisi, Michel
  Ntemgwa, Claudine Matte, Jean-Guy Baril, R{\'e}jean Thomas, Danielle Rouleau,
  Julie Bruneau, et~al.
\newblock High rates of forward transmission events after acute/early {HIV-1}
  infection.
\newblock {\em Journal of Infectious Diseases}, 195(7):951--959, 2007.

\bibitem{buve2002spread}
Anne Buv{\'e}, Kizito Bishikwabo-Nsarhaza, and Gladys Mutangadura.
\newblock The spread and effect of hiv-1 infection in sub-saharan africa.
\newblock {\em The Lancet}, 359(9322):2011--2017, 2002.

\bibitem{chibo:acute}
Doris Chibo, Matthew Kaye, and Chris Birch.
\newblock {HIV} transmissions during seroconversion contribute significantly to
  new infections in men who have sex with men in {Australia}.
\newblock {\em AIDS research and human retroviruses}, 28(5):460--464, 2012.

\bibitem{decreusefond:volz_limit}
Laurent Decreusefond, Jean-St\'{e}phane Dhersin, Pascal Moyal, and Viet~Chi
  Tran.
\newblock Large graph limit for an {SIR} process in random network with
  heterogeneous connectivity.
\newblock {\em The Annals of Applied Probability}, 22(2):541--575, 2012.

\bibitem{eaton2011concurrent}
Jeffrey~W Eaton, Timothy~B Hallett, and Geoffrey~P Garnett.
\newblock Concurrent sexual partnerships and primary hiv infection: a critical
  interaction.
\newblock {\em AIDS and Behavior}, 15(4):687--692, 2011.

\bibitem{epstein:concurrent}
H.~Epstein.
\newblock The mathematics of concurrent partnerships and {HIV}: A commentary on
  {L}urie and {R}osenthal, 2009.
\newblock {\em AIDS and Behavior}, 14(1):29--30, 2010.

\bibitem{epstein2011concurrent}
Helen Epstein and Martina Morris.
\newblock Concurrent partnerships and hiv: an inconvenient truth.
\newblock {\em Journal of the International AIDS Society}, 14(1):13, 2011.

\bibitem{goodreau2011decade}
Steven~M Goodreau.
\newblock A decade of modelling research yields considerable evidence for the
  importance of concurrency: a response to sawers and stillwaggon.
\newblock {\em Journal of the International AIDS Society}, 14(1):12, 2011.

\bibitem{janson:SIRproof}
Svante Janson, Malwina Luczak, and Peter Windridge.
\newblock Law of large numbers for the {SIR} epidemic on a random graph with
  given degrees.
\newblock {\em Random Structures \& Algorithms}, 45(4):724--761, 2014.

\bibitem{kalipeni2004hiv}
Ezekiel Kalipeni, Susan Craddock, Joseph~R Oppong, and Jayati Ghosh, editors.
\newblock {\em HIV and AIDS in Africa: beyond epidemiology.}
\newblock Malden Massachusetts Blackwell Publishing 2004., 2004.

\bibitem{kamp2010untangling}
Christel Kamp.
\newblock Untangling the interplay between epidemic spread and transmission
  network dynamics.
\newblock {\em PLoS Comput Biol}, 6(11):e1000984, 2010.

\bibitem{karrer:message}
Brian Karrer and Mark E.~J. Newman.
\newblock Message passing approach for general epidemic models.
\newblock {\em Physical Review E}, 82:016101, 2010.

\bibitem{kiss:EoN}
Istvan~Z Kiss, Joel~C Miller, and P{\'e}ter~L Simon.
\newblock {\em Mathematics of epidemics on networks: from exact to approximate
  models}.
\newblock IAM. Springer, 2017.

\bibitem{kretzschmar:concurrent}
M.~Kretzschmar, R.G. White, and M.~Cara{\"e}l.
\newblock Concurrency is more complex than it seems.
\newblock {\em AIDS (London, England)}, 24(2):313, 2010.

\bibitem{leung2016dangerous}
Ka~Yin Leung and Odo Diekmann.
\newblock Dangerous connections: on binding site models of infectious disease
  dynamics.
\newblock {\em Journal of Mathematical Biology}, pages 1--53, 2016.

\bibitem{leung2015concurrency}
Ka~Yin Leung and Mirjam Kretzschmar.
\newblock Concurrency can drive an hiv epidemic by moving $r_0$ across the
  epidemic threshold.
\newblock {\em Aids}, 29(9):1097--1103, 2015.

\bibitem{leung:demographic}
Ka~Yin Leung, Mirjam Kretzschmar, and Odo Diekmann.
\newblock Dynamic concurrent partnership networks incorporating demography.
\newblock {\em Theoretical population biology}, 82(3):229--239, 2012.

\bibitem{leung:disease}
Ka~Yin Leung, Mirjam Kretzschmar, and Odo Diekmann.
\newblock $si$ infection on a dynamic partnership network: characterization of
  $r_0$.
\newblock {\em Journal of mathematical biology}, 71(1):1--56, 2014.

\bibitem{liljeros}
Fredrik Liljeros, Christofer~R. Edling, Lu\'{\i}s A.~Nunes Amaral, H.~Eugene
  Stanley, and Yvonne {\AA}berg.
\newblock The web of human sexual contacts.
\newblock {\em Nature}, 411(6840):907--908, 2001.

\bibitem{lloyd2004frequency}
James~O Lloyd-Smith, Wayne~M Getz, and Hans~V Westerhoff.
\newblock Frequency--dependent incidence in models of sexually transmitted
  diseases: portrayal of pair--based transmission and effects of illness on
  contact behaviour.
\newblock {\em Proceedings of the Royal Society of London B: Biological
  Sciences}, 271(1539):625--634, 2004.

\bibitem{lurie:concurrent2}
M.N. Lurie and S.~Rosenthal.
\newblock The concurrency hypothesis in sub-{Saharan} {Africa}: Convincing
  empirical evidence is still lacking. {Response} to {Mah} and {Halperin},
  {Epstein}, and {Morris}.
\newblock {\em AIDS and Behavior}, 14(1):34--37, 2010.

\bibitem{lurie:concurrent}
M.N. Lurie and S.~Rosenthal.
\newblock Concurrent partnerships as a driver of the {HIV} epidemic in
  sub-{Saharan} {Africa}? {The} evidence is limited.
\newblock {\em AIDS and Behavior}, 14(1):17--24, 2010.

\bibitem{mah:concurrent}
T.L. Mah and D.T. Halperin.
\newblock Concurrent sexual partnerships and the {HIV} epidemics in {Africa}:
  evidence to move forward.
\newblock {\em AIDS and Behavior}, 14(1):11--16, 2010.

\bibitem{mah:evidence}
T.L. Mah and D.T. Halperin.
\newblock The evidence for the role of concurrent partnerships in {Africa}'s
  {HIV} epidemics: a response to {L}urie and {R}osenthal.
\newblock {\em AIDS and Behavior}, 14(1):25--28, 2010.

\bibitem{miller:volz}
Joel~C. Miller.
\newblock A note on a paper by {Erik Volz}: {SIR} dynamics in random networks.
\newblock {\em Journal of Mathematical Biology}, 62(3):349--358, 2011.

\bibitem{miller:ebcm_overview}
Joel~C. Miller, Anja~C. Slim, and Erik~M. Volz.
\newblock Edge-based compartmental modelling for infectious disease spread.
\newblock {\em Journal of the Royal Society Interface}, 9(70):890--906, 2012.

\bibitem{miller:ebcm_structure}
Joel~C. Miller and Erik~M. Volz.
\newblock Incorporating disease and population structure into models of {SIR}
  disease in contact networks.
\newblock {\em PLoS ONE}, 8(8):e69162, 2013.

\bibitem{morris:barking}
M.~Morris.
\newblock Barking up the wrong evidence tree. {Comment} on {L}urie \&
  {R}osenthal, ``{C}oncurrent partnerships as a driver of the {HIV} epidemic in
  sub-{S}aharan {A}frica? {T}he evidence is limited''.
\newblock {\em AIDS and Behavior}, 14(1):31--33, 2010.

\bibitem{morris:concurrentnetwork}
Martina Morris and Mirjam Kretzschmar.
\newblock Concurrent partnerships and transmission dynamics in networks.
\newblock {\em Social Networks}, 17(3):299--318, 1995.

\bibitem{morris:concurrent}
Martina Morris and Mirjam Kretzschmar.
\newblock Concurrent partnerships and the spread of {HIV}.
\newblock {\em AIDS}, 11(5):641--648, 1997.

\bibitem{murray2014global}
Christopher~JL Murray, Katrina~F Ortblad, Caterina Guinovart, Stephen~S Lim,
  Timothy~M Wolock, D~Allen Roberts, Emily~A Dansereau, Nicholas Graetz, Ryan~M
  Barber, Jonathan~C Brown, et~al.
\newblock Global, regional, and national incidence and mortality for hiv,
  tuberculosis, and malaria during 1990--2013: a systematic analysis for the
  global burden of disease study 2013.
\newblock {\em The Lancet}, 384(9947):1005--1070, 2014.

\bibitem{newman:structurereview}
Mark E.~J. Newman.
\newblock The structure and function of complex networks.
\newblock {\em SIAM Review}, 45(2):167--256, 2003.

\bibitem{pastor2015epidemic}
Romualdo Pastor-Satorras, Claudio Castellano, Piet Van~Mieghem, and Alessandro
  Vespignani.
\newblock Epidemic processes in complex networks.
\newblock {\em Reviews of modern physics}, 87(3):925, 2015.

\bibitem{pilcher:acute_hiv}
C.D. Pilcher, H.C. Tien, J.J. Eron, P.L. Vernazza, S.Y. Leu, P.W. Stewart, L.E.
  Goh, and M.S. Cohen.
\newblock Brief but efficient: acute {HIV} infection and the sexual
  transmission of {HIV}.
\newblock {\em Journal of Infectious Diseases}, 189(10):1785, 2004.

\bibitem{pilcher2007amplified}
Christopher~D Pilcher, George Joaki, Irving~F Hoffman, Francis~EA Martinson,
  Clement Mapanje, Paul~W Stewart, Kimberly~A Powers, Shannon Galvin, David
  Chilongozi, Syze Gama, et~al.
\newblock Amplified transmission of hiv-1: comparison of hiv-1 concentrations
  in semen and blood during acute and chronic infection.
\newblock {\em AIDS (London, England)}, 21(13):1723, 2007.

\bibitem{pinkerton2008probability}
Steven~D Pinkerton.
\newblock Probability of hiv transmission during acute infection in rakai,
  uganda.
\newblock {\em AIDS and Behavior}, 12(5):677--684, 2008.

\bibitem{sawers2010concurrent}
Larry Sawers and Eileen Stillwaggon.
\newblock Concurrent sexual partnerships do not explain the hiv epidemics in
  africa: a systematic review of the evidence.
\newblock {\em Journal of the International AIDS society}, 13(1):34, 2010.

\bibitem{volz:cts_time}
Erik~M. Volz.
\newblock {SIR} dynamics in random networks with heterogeneous connectivity.
\newblock {\em Journal of Mathematical Biology}, 56(3):293--310, 2008.

\bibitem{volz:dynamic_network}
Erik~M. Volz and Lauren~Ancel Meyers.
\newblock Susceptible--infected--recovered epidemics in dynamic contact
  networks.
\newblock {\em Proceedings of the Royal Society B: Biological Sciences},
  274(1628):2925--2933, 2007.

\bibitem{wallinga2007generation}
Jacco Wallinga and Marc Lipsitch.
\newblock How generation intervals shape the relationship between growth rates
  and reproductive numbers.
\newblock {\em Proceedings of the Royal Society of London B: Biological
  Sciences}, 274(1609):599--604, 2007.

\bibitem{wang2017unification}
Wei Wang, Ming Tang, H~Eugene Stanley, and Lidia~A Braunstein.
\newblock Unification of theoretical approaches for epidemic spreading on
  complex networks.
\newblock {\em Reports on Progress in Physics}, 80(3):036603, 2017.

\end{thebibliography}

\end{document}